\documentclass{article}

\usepackage[many]{tcolorbox}
\usepackage{mdframed}
\usepackage{enumitem}
\usepackage{subfigure}
\usepackage{wrapfig}
\usepackage{twemojis}

    \usepackage[final]{neurips_data_2024}

\usepackage{amsmath}
\usepackage{amssymb}
\usepackage{mathtools}
\usepackage{amsthm}
\usepackage{balance}

\usepackage[utf8]{inputenc} 
\usepackage[T1]{fontenc}    
\usepackage{hyperref}       
\usepackage{url}            
\usepackage{booktabs}       
\usepackage{amsfonts}       
\usepackage{nicefrac}       
\usepackage{microtype}      
\usepackage{xcolor}         

\usepackage[capitalize,noabbrev]{cleveref}

\theoremstyle{plain}

\theoremstyle{definition}

\theoremstyle{remark}

\usepackage[textsize=tiny]{todonotes}

\usepackage{lly}

\newtcolorbox{generalpretrainbar}[2][]
{
width=#2,right=0mm,height=0.35cm,size=minimal,valign=center,top=0pt,bottom=0pt,before=\noindent,after skip=0pt,
colback=barblue1,
}

\newtcolorbox{generaltunebar}[2][]
{
width=#2,right=0mm,height=0.35cm,size=minimal,valign=center,top=0pt,bottom=0pt,before=\noindent,after=\noindent,after skip=0pt,
colback=barblue2,
}

\newtcolorbox{generalpretrainbarinline}[2][]
{
width=#2,right=0mm,height=0.35cm,size=minimal,valign=center,top=0pt,bottom=0pt,before=\noindent,after skip=0pt,after=\noindent,
colback=barblue1,
}

\newtcolorbox{generaltunebarinline}[2][]
{
width=#2,right=0mm,height=0.35cm,size=minimal,valign=center,top=0pt,bottom=0pt,before=\noindent,after=\noindent,after skip=0pt,after=\noindent,
colback=barblue2,
}

\newtcolorbox{codepretrainbar}[2][]
{
width=#2,right=0mm,height=0.35cm,size=minimal,valign=center,top=0pt,bottom=0pt,before=\noindent,after skip=0pt,
colback=barorange1,
}

\newtcolorbox{codetunebar}[2][]
{
width=#2,right=0mm,height=0.35cm,size=minimal,valign=center,top=0pt,bottom=0pt,before=\noindent,after=\noindent,after skip=0pt,
colback=barorange2,
}

\newtcolorbox{defaultbar}[2][]
{
width=#2,right=0mm,height=0.35cm,size=minimal,valign=center,top=0pt,bottom=0pt,before=\noindent,after=\noindent,after skip=0pt,
colback=gray!20,
}

\newtcolorbox{codepretrainbarinline}[2][]
{
width=#2,right=0mm,height=0.35cm,size=minimal,valign=center,top=0pt,bottom=0pt,before=\noindent,after skip=0pt,after=\noindent,
colback=barorange1,
}

\newtcolorbox{codetunebarinline}[2][]
{
width=#2,right=0mm,height=0.35cm,size=minimal,valign=center,top=0pt,bottom=0pt,before=\noindent,after=\noindent,after skip=0pt,after=\noindent,
colback=barorange2,
}

\newtcolorbox{pretrainbar}[2][]
{
width=#2,right=0mm,height=0.35cm,size=minimal,valign=center,top=0pt,bottom=0pt,before=\noindent,after skip=0pt,
colback=bargreen1,
}

\newtcolorbox{tunebar}[2][]
{
width=#2,right=0mm,height=0.35cm,size=minimal,valign=center,top=0pt,bottom=0pt,before=\noindent,after=\noindent,after skip=0pt,
colback=bargreen2,
}

\newtcolorbox{pretrainbarinline}[2][]
{
width=#2,right=0mm,height=0.35cm,size=minimal,valign=center,top=0pt,bottom=0pt,before=\noindent,after skip=0pt,after=\noindent,
colback=bargreen1,
}

\newtcolorbox{tunebarinline}[2][]
{
width=#2,right=0mm,height=0.35cm,size=minimal,valign=center,top=0pt,bottom=0pt,before=\noindent,after=\noindent,after skip=0pt,after=\noindent,
colback=bargreen2,
}

\title{\sysname: Evaluating the Question-Answering Capabilities of Code Large Language Models}

\author{%
    Linyi Li\\
    Simon Fraser University\\
    \texttt{\href{mailto:linyi_li@illinois.edu}{linyi\_li@sfu.ca}} \\
    \And
    Shijie Geng\\
    ByteDance Inc \& Rutgers University\\
    \texttt{\href{mailto:sg1309@rutgers.edu}{sg1309@rutgers.edu}}
    \And
    Zhenwen Li\thanks{Equal contribution.}\quad Yibo He\samethanks\quad Hao Yu\samethanks\quad Ziyue Hua\samethanks\\
    Peking University\\
    \And
    Guanghan Ning\\
    ByteDance Inc\\
    \And
    Siwei Wang\\
    ByteDance Inc\\
    \And
    Tao Xie\\
    Key Lab of HCST (PKU), MOE\\
    \texttt{\href{mailto:taoxie@pku.edu.cn}{taoxie@pku.edu.cn}}
    \And
    Hongxia Yang\\
    The Hong Kong Polytechnic University (PolyU)\\
    \texttt{\href{mailto:hongxia.yang@polyu.edu.hk}{hongxia.yang@polyu.edu.hk}} \\
}

\begin{document}

\maketitle

\begin{abstract}

Large Language Models for code~(code LLMs) have witnessed tremendous progress in recent years.
With the rapid development of code LLMs,
many popular evaluation benchmarks, such as HumanEval, DS-1000, and MBPP, have emerged to measure the performance of code LLMs with a particular focus on code generation tasks.
However, they are insufficient to cover the full range of expected capabilities of code LLMs, which span beyond code generation to answering diverse coding-related questions.
To fill this gap, we propose \textbf{\sysname}, the \textbf{first large-scale freeform question-answering~(QA) benchmark for code} to our knowledge, comprising 234 carefully selected high-quality Stack Overflow questions that span across 15 programming languages.
\sysname uses four types of model-free automatic metrics to evaluate response correctness where domain experts carefully concretize the criterion for each question.
We conduct a systematic evaluation for over 100 latest code LLMs on \sysname, leading to a series of novel and insightful findings. 
Our detailed analyses showcase potential directions for further advancement of code LLMs.
\sysname is fully open source at \url{https://infi-coder.github.io/infibench} and continuously expanding to foster more scientific and systematic practices for code LLM evaluation.
\end{abstract}

\begin{figure}[htbp]
    \includegraphics[width=\linewidth]{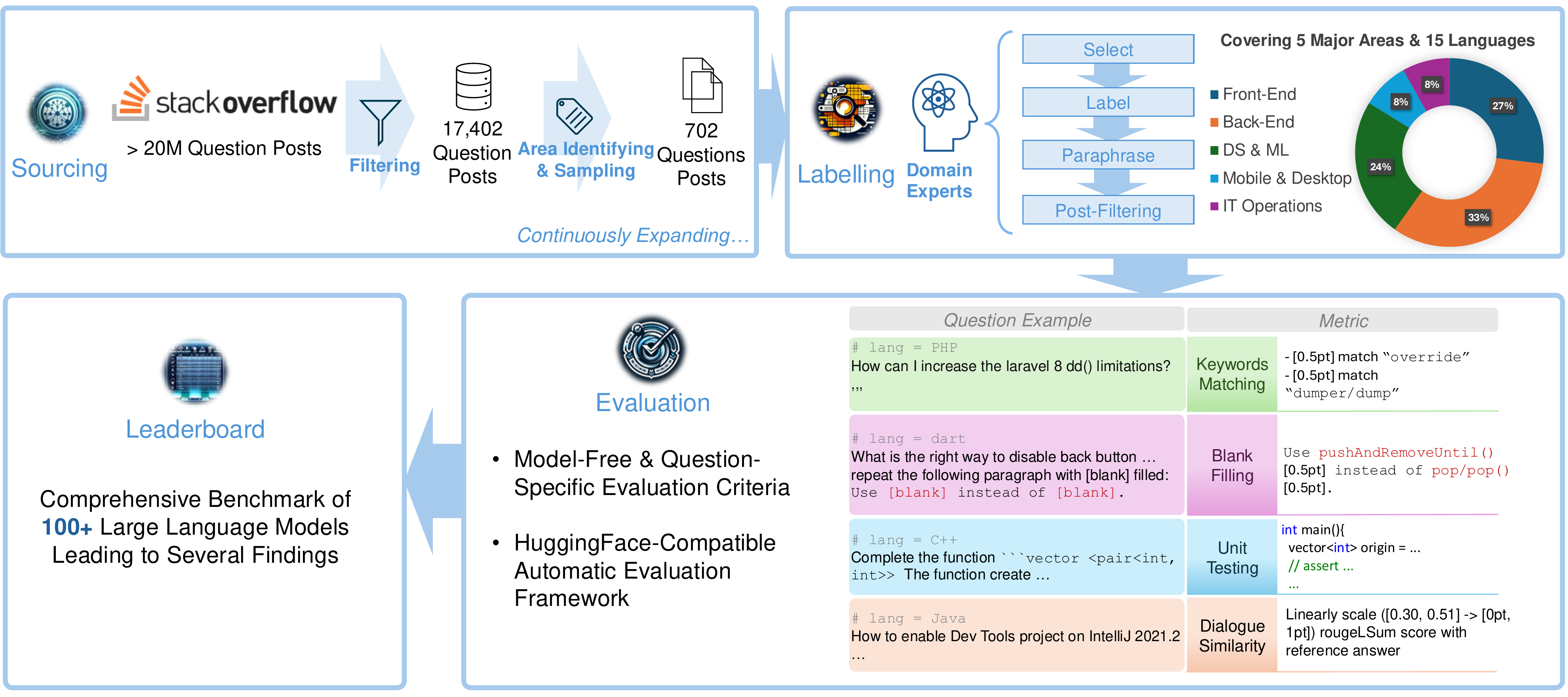}
    \vspace{-1em}
    \caption{\small \textbf{\sysname overview}. We construct the \sysname benchmark by filtering high-quality and diverse question posts from Stack Overflow and annotating question-level evaluation criteria with domain experts. With an model-free automatic evaluation framework, we evaluate over 100 latest code LLMs~(one of the most extensive evaluations for code LLMs to the best of our knowledge), leading to several insightful findings.}
    \label{fig:example}
    \vspace{-0.5em}
\end{figure}

\section{Introduction}
\label{sec:intro}
In recent years, Large Language Models~(LLMs) have been revolutionizing the software development landscape~\citep{hou2023large,fan2023large}, demonstrating exceedingly strong and comprehensive capabilities in comprehending, generating, debugging, and summarizing code~\citep{chen2021evaluating,li2022competition}. For example, code LLM-powered products like GitHub Copilot~\citep{github2023copilot} reached millions of active users within just one year of their launch.


Alongside the huge success of proprietary LLMs such as GPT-3.5 / GPT-4~\citep{OpenAI2023GPT4TR} and Gemini~\citep{team2023gemini}, the development 
of open-source code LLMs\footnote{We define code LLMs as LLMs that show decent capabilities in the code domain, no matter whether they are exclusively trained or finetuned with code data or not.}~\citep{nijkamp2023codegen,touvron2023llama,roziere2023code,luo2023wizardcoder} has been advancing at an unprecedented fast pace. As of June 2024, the Hugging Face Open LLM Leaderboard~\citep{open-llm-leaderboard} has cataloged over 3,300 submissions of such models.

\begin{wrapfigure}{l}{0.45\textwidth}
    \centering
    \includegraphics[width=\linewidth]{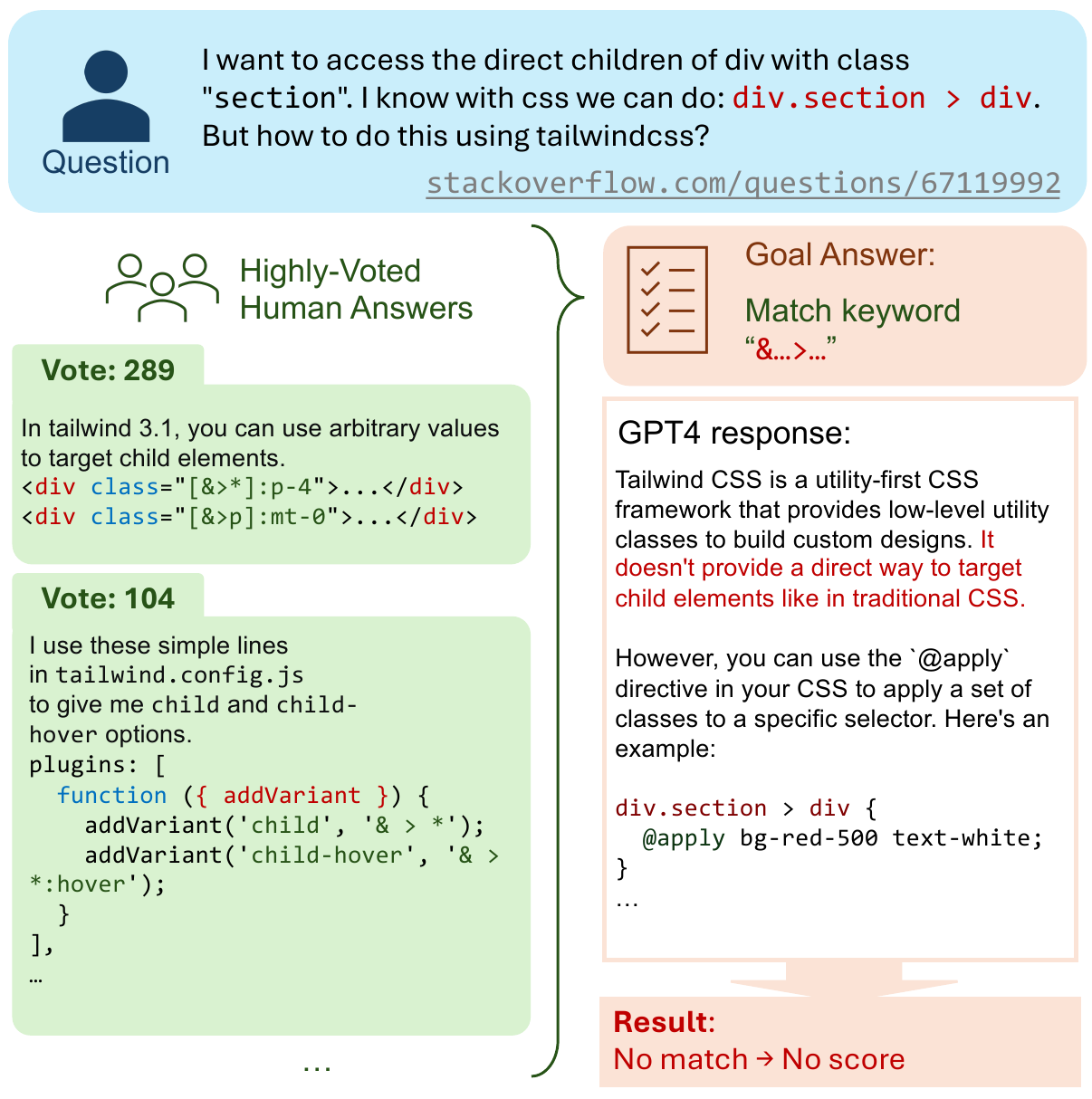}
    \vspace{-0.8em}
    \caption{\small A challenging question paraphrased from Stack Overflow where GPT-4 fails to answer.}
    \label{fig:challenging-scenario}
    \vspace{-1em}
\end{wrapfigure}

Given the plethora of code LLMs available, the development of reliable code benchmarks seems to lag in four aspects:
(1)~\textbf{Benchmarks beyond code generation are relatively few.}
Benchmarks for code LLMs typically focus on a specific task or domain, often overly focus on code generation. 
For example, the widely-used HumanEval~\citep{chen2021evaluating} and MBPP~\citep{austin2021program} purely focus on Python code generation, and DS-1000~\citep{lai2023ds} focuses on Python code generation in the field of data science.
(2)~\textbf{Independent code benchmarks are relatively few.}
Recent efforts evolve existing benchmarks~(e.g., HumanEval) to include more scenarios~\citep{muennighoff2023octopack}, languages~\citep{zheng2023codegeex}, and tests~\citep{liu2023is}.
However, these efforts lead to a series of benchmarks sharing the same source data~(e.g., HumanEval Python problems), reducing score independence.
(3)~\textbf{Existing code benchmarks are saturating.}
Strong LLMs are saturating existing benchmarks, e.g., GPT-4 has already achieved 90.2\% Pass@1 score on HumanEval~\citep{HelloGPT4o}, while in real-world scenarios, GPT-4 can still fail as exemplified in \Cref{fig:challenging-scenario}.
(4)~\textbf{Common benchmarks may be contaminated.}
Some LLMs have unconventional high performance in common benchmarks and are suspected to have memorized benchmark-related data~\citep{dekoninck2024constat,xu2024benchmarking,matton2024leakage}, obscuring the evaluation results.
\textit{Can we systematically and comprehensively evaluate code LLMs' abilities in challenging real-world usage scenarios?}



        \begin{table*}[!t]
            \centering
            \caption{\small \textbf{Comparison between \sysname and common existing benchmarks.} Existing benchmarks weigh heavily on code generation, unit-test-based evaluation, and major programming languages. \sysname processes a much higher diversity to reflect real-world code LLMs' usage scenarios. More discussion in \Cref{subsec:comparison-with-existing-bench}.}
            \centering
            \resizebox{\linewidth}{!}{
            \begin{tabular}{cccccc}
                \toprule
                Benchmark & Domain & \# Question & Evaluation & Data Source & Highest LLM Score \\
                \midrule
                HumanEval~\citep{chen2021evaluating} & Python Programming & 164 & Test Cases & Hand-Written & 90.2\% \\
                MBPP~\citep{austin2021program} & Python Programming & 974 & Test Cases & Hand-Written & 81.1\% \\
                APPS~\citep{hendrycks2021measuring} & Python Programming & 10,000 & Test Cases & Competitions & / (no report yet)  \\
                DS-1000~\citep{lai2023ds} & Python Programming & 1,000 & Test Cases + Surface Form Constraints & StackOverflow & / (no report yet) \\
                HumanEval+~\citep{liu2023is} & Python Programming & 164 & Augmented Test Cases & HumanEval & 86.6\% \\
                HumanEvalPack~\citep{muennighoff2023octopack} & Repair, Explain, Generation in 6 Languages & 2,952 & Test Cases & HumanEval & 47.8\%/52.1\%/78.3\% \\
                LBPP~\citep{matton2024leakage} & Python Programming & 161 & Test Cases & Hand-Written & 64\% \\
                SWE-bench~\citep{jimenez2024swe} & Python Debugging / Repair  & 2,294 & Test Cases & GitHub & 22.06\% \\
                SWE-bench Verified~\citep{openai2024sweverified} & Python Debugging / Repair & 500 & Test Cases & SWE-bench & 45.20\% \\
                \hline
                \multirow{2}{*}{\sysname} & Free-Form Code Question & \multirow{2}{*}{234} & Keyword + Blank Filling +  & Stack  & \multirow{2}{*}{70.64\%} \\
                & Answering in 15 Languages & & Test Cases + Text Similarity & Overflow  & \\
                \bottomrule
            \end{tabular}
            }
            \label{tab:benchmark-comparison}
            \vspace{-1em}
        \end{table*}


To answer the question, we introduce \sysname, a systematic benchmark for evaluating the free-form question-answering capabilities of code LLMs.
As the first benchmark of its kind,
the core principle of \sysname aims to accurately represent how developers interact with and utilize such models in real-world scenarios.
To achieve this, \sysname comprises 234 questions
that are carefully selected and proportionally filtered from the natural high-quality question distribution of Stack Overflow, without any constraints on topics, programming languages, question types, or answer forms.
As a result, the curated 234 questions span 15 programming languages and 5 major areas: \textit{front-end}, \textit{back-end}, \textit{DS\&ML~(data science and machine learning)}, \textit{mobile and desktop}, and \textit{ITOps~(information technology operations)}.

Question diversity comes with evaluation challenges for two reasons.
(1) Lack of metric. Unlike code generation or multiple-choice benchmarks, which can be evaluated through standardized methods like unit testing, 
there is no universal metric for response correctness for free-form questions.
(2) Challenges with model-based evaluation. Model-based evaluations such as those involving GPT-4 are not only costly but also raise concerns about privacy and bias.

To mitigate the evaluation challenges, \sysname includes an automatic evaluation framework that integrates four types of \emph{model-free} metrics: keyword matching, blank filling, unit testing, and dialogue similarity. 
For each question, we invite industry domain experts to paraphrase the prompt, select the most appropriate metric, and write down the concrete criteria using domain-specific knowledge, with highly-voted answers from Stack Overflow as a reference.
These questions and evaluation criteria are then cross-validated to ensure correctness and objectiveness and further calibrated to improve consistency across languages.
Human experiments show that \sysname evaluation aligns with humans better than LLM-based evaluation, achieving 85.1\% agreement rate compared to 77.8\% achieved by GPT-4o-based evaluation.

As a novel and systematic benchmark disjoint with existing ones in terms of both forms and data sources, we believe that \sysname is an ideal tool to measure existing code LLMs objectively.
Hence, we conduct a systematic evaluation for \textbf{over 100 code LLMs} spanning both proprietary and open-source worlds using the \sysname framework --- the latest and most extensive evaluation for code LLMs to the best of our knowledge.
Our evaluation leads to several insightful findings:
(1)~On \sysname, GPT-4 achieves a score of $70.64\%$, being far from perfect but still far exceeding the most capable open-source models as of June 2024. On the other hand, GPT3.5 is surpassed by a few open-source models.
(2)~At similar model sizes, coding LLMs are usually visibly stronger than general LLMs; finetuning LLMs are usually visibly stronger than base LLMs.
(3)~The performance differences between different model families are huge, where one model could surpass another with less than 1/10 parameters, highlighting the importance of training data quality and techniques.
(4)~The scaling law is empirically verified for open-source models with fewer than 40B parameters, but not for those with more, where a turning point emerges. 
%
\sysname is fully open source under CC BY-SA 4.0 license and continuously expanding\footnote{In other words, \textbf{infi}nitely expanding, after which the \textbf{bench}mark is named.}, including both the benchmark and Hugging-Face-compatible evaluation tools.
All resources are available at {\url{https://infi-coder.github.io/infibench}}.



\section{Benchmark Creation}
\label{sec:creation}

\sysname is created from a high-quality subset of Stack Overflow questions up until June 14, 2023. In this section, we describe the data curation process and the evaluation framework in detail.

    \subsection{Data Curation}
        \label{subsec:data-curation}

Stack Overflow is a question-and-answer website for developers with more than 24 million registered users as of June 2024~\citep{StackExchangeStatistics}.
Since the website is a large collection of natural and diverse coding questions from real-world developers, we believe that questions from Stack Overflow can effectively evaluate code LLM's capabilities in real-world usage scenarios.

The full Stack Overflow dataset contains 23.54 million question posts and 34.68 million answer posts.
Each question post has a total view count. Each answer post is attached to a question and has a vote count.
The question creator can choose one answer as officially accepted.

\begin{wrapfigure}{r}{0.45\textwidth}
            \vspace{-2.3em}
            \centering
            \includegraphics[width=\linewidth]{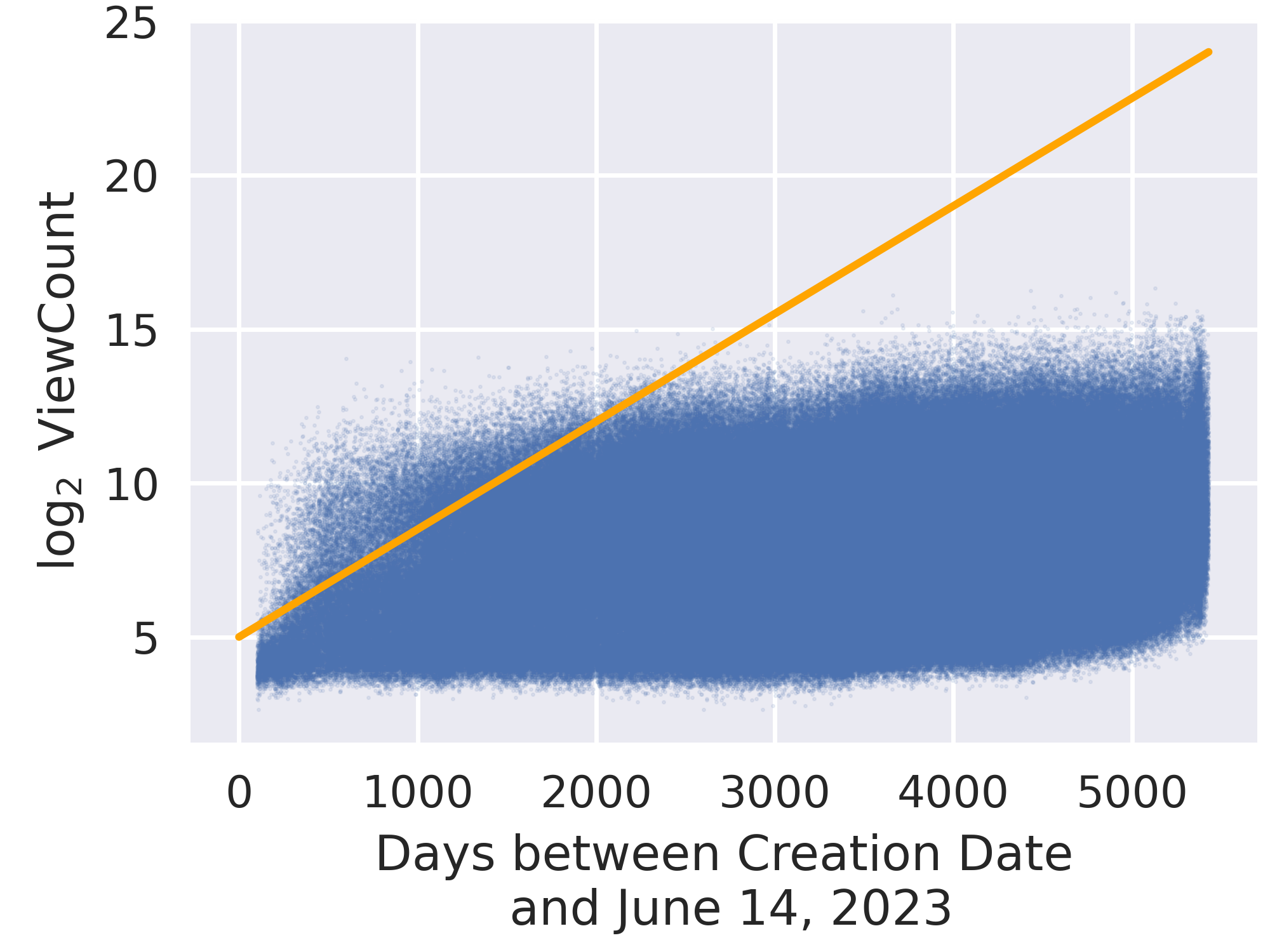}
            \vspace{-1.5em}
            \caption{\small Scatter plot of filtered Stack Overflow questions. Questions above the orange line kept.}
            \label{fig:date-view-plot}
            \vspace{-1em}
        \end{wrapfigure}

As we aim to create a benchmark where the correctness evaluation criteria are clear, we view the positively voted answers as an important reference source.
Hence, we choose to keep only the questions that have at least three positively voted answers and an officially accepted answer, which turn out to be 1,090,238 questions.
For these one million questions, we choose to keep questions that are frequently viewed and relatively new. To fulfill this criterion, we draw a scatter plot of these $\approx$ 1 million questions, plotting the number of days since their creation until June 14, 2023~(data collection end-date) on the $x$-axis against the logarithm of their view counts on the $y$-axis. As shown in \Cref{fig:date-view-plot}, we empirically determine to keep questions that lie above the line connecting $(0,5)$ and $(3000, 15.5)$, resulting in a subset of 17,402 questions.

Utilizing the mandatory question tags of these questions, we then manually construct a tag tree that covers the 200 most frequent tags, enabling us to identify the top programming languages and areas for 14,330 out of these 17,402 questions. These questions are from 24 programming languages, with each language being categorized into one primary area among the five~(front-end, back-end, DS\&ML, mobile and desktop, and ITOps). Lastly, we exclude 6 programming languages that either describe data or are domain-specific: JSON, regex, Markdown, YAML, CSV, and SQL.
As a result, we compile 13,854 questions that serve as the \emph{initial seed set}.
        
    \subsection{Sampling}

        \label{subsec:sampling}
    
Based on a user study of developers' demand from our organization,
we allocate the tentative area quota to be 25\%, 25\%, 25\%, 15\%, and 10\% for front-end, back-end, DS\&ML, mobile and desktop, and IT Ops, respectively.
Inspired by HumanEval size and considering the labelling labor cost, we set 200 questions as the target benchmark size.
Hence, the tentative size quotas by area are 50, 50, 50, 30, and 20 respectively.
We then proportionally distribute the area quotas to language quotas based on the frequency of each language in the initial seed set.
However, we observe that following this rule, certain languages such as CSS and C/C++ end up with fewer than 10 questions, which may yield unreliable language-level sub-score, so, for these languages, we set their quotas to 10.

As a result, we derive the \emph{tentative} question quota for each language as shown in \Cref{tab:quota}, which sums up to 270 
questions.
After determining the tentative question quota, we uniformly sample from the initial seed set a roughly two times larger pool for the domain expects to select and annotate. 

        \begin{table}[!t]
            \centering
            \caption{\small \textbf{\sysname data statistics by area and language}. We uniformly sample a subset from the initial seed set~(see \Cref{subsec:data-curation}) according to the area quota~(see \Cref{subsec:sampling}) for domain experts to select questions and annotate the correctness criterion to construct the benchmark.}
            \resizebox{0.9\linewidth}{!}{
            \begin{tabular}{c|c|cc|c|cccc}
                \toprule
                \multirow{3}{*}{Area} & \multirow{3}{*}{Language} & \multicolumn{2}{c|}{Initial Seed Set} & Tentative & \multicolumn{4}{c}{Final \sysname Benchmark}  \\
                & & \multirow{2}{*}{\# Questions} & \% Area & \# Questions & \# Questions & \% Questions & \# Area & \% Area \\
                & & & Quota & Quota & Quota & Quota & Quota & Quota \\
                \midrule
                \multirow{3}{*}{Front-End} & Javascript & 4912 & \multirow{3}{*}{40.41\%}  & 44 & 44 & 18.80\% & \multirow{3}{*}{63} & \multirow{3}{*}{26.92\%} \\
                & CSS & 87 & & 10 & 10 & 4.27\% & & \\
                & HTML & 600 & & 10 & 9 & 3.85\% & & \\
                \hline
                \multirow{7}{*}{Back-End} & Java & 930  & \multirow{7}{*}{18.71\%} & 18 & 17 & 7.26\% & \multirow{7}{*}{77} & \multirow{7}{*}{32.91\%} \\
                & C\# & 629  & &12 & 12 & 5.13\% & & \\
                & PHP & 462  & & 10 & 9 & 3.85\% & & \\
                & Go & 117  & & 10 & 9 &  3.85\% & & \\
                & Ruby & 71  & & 10 & 10 & 4.27\% & & \\
                & Rust & 96  & & 10 & 10 & 4.27\% & & \\
                & C/C++ & 287  & & 10 & 10 & 4.27\% & & \\
                \hline
                \multirow{2}{*}{DS \& ML} & Python & 2779  & \multirow{2}{*}{21.39\%} & 47 & 47 & 20.09\% & \multirow{2}{*}{56} & \multirow{2}{*}{23.93\%}  \\
                & R & 184 & & 10 & 9 & 3.85\% & & \\
                \hline
                \multirow{4}{*}{\shortstack[c]{Mobile \& \\ Desktop}} & Dart & 1562  & \multirow{4}{*}{18.13\%} & 19 & 19 & 8.12\% & 19 & 8.12\% \\
                & Kotlin & 383  & & 10 & \multicolumn{4}{c}{\multirow{3}{*}{Removed during Post-Filtering~(see \Cref{subsec:annotation})}} \\
                & Swift & 551  & & 10 &  & & & \\
                & VBA  & 16 & & 9 & & & & \\
                \hline
                IT Ops. & Bash & 188 & 1.36\% & 21 & 19 & 8.12\% & 19 & 8.12\%  \\
                \hline\hline
                \multicolumn{2}{c|}{Total} & 13854 & 100.0\% & 270 & 234 & 100.00 \% & 234 & 100.00\% \\
                \bottomrule
            \end{tabular}
            }
            \label{tab:quota}
        \end{table}

    \vspace{-0.5em}
    \subsection{Human Annotation}
    \vspace{-0.5em}

        \label{subsec:annotation}

        We recruited five domain experts inside our company to create the benchmark, each in charge of one area.
        The annotation process is composed of three steps:
        \begin{itemize}[leftmargin=*,noitemsep]
            \item \textbf{Step 1: Question Selection and Type Annotation.} Domain experts select high-quality questions from the inspecting set and annotate the question type to be one of the four: code completion, code debugging, config and environment debugging, and knowledge question-answering.

            \item \textbf{Step 2: Prompt Paraphrasing.} Domain experts paraphrase and simplify the original question body into succinct and explicit instructions. We include this step for two main purposes: 
            (1) Reduce domain gap. From user-shared conversations collected from ShareGPT, we observe that when interacting with code LLMs, users tend to provide short and direct instructions like ``Fix problem...'' and ``Debug code...''. However, when posting Stack Overflow questions, users tend to be lengthy with courtesy words.
            We ask the domain experts to paraphrase the question to code LLM user's style without changing the semantics.
            (2) Reduce the impact of memorization and  data contamination. Some code LLMs may be trained or finetuned with Stack Overflow data. Paraphrasing the questions can help to mitigate the result advantages of these models. Benchmark results in \Cref{tab:main_abstract_barplot} reveal the effectiveness of this step where copying Stack Overflow answers only achieves a 65.18\% score.
            We defer further discussion in \Cref{subsec:data-contamination}.

            \item \textbf{Step 3: Correctness Criterion Annotation.} 
            Domain experts choose one or multiple evaluation metrics from our supported ones~(see \Cref{subsec:evaluation-framework-and-score-computing}) and annotate the concrete criterion following a YAML schema. External files can be attached if needed, e.g., unit tests and reference answers.
        \end{itemize}

        \textbf{Calibration and Post-Filtering.}
        To improve annotation consistency and objectiveness, we introduce a few checkpoints for domain experts to read others' annotated cases, discuss them, and reach consensus for controversial cases.
        %
        After the 270 tentative questions were annotated, we then ran an initial evaluation of all these questions on over 30 code LLMs.
        This initial evaluation helps us to identify questions whose criteria are incorrect or out of distribution.
        We filter out these questions and then remove all questions from Kotlin, Swift, and VBA languages since the questions in these languages are too few after filtering.
        After this calibration and post-filtering process, the final benchmark includes 234 questions spanning over 15 languages.
        Their statistics are shown in \Cref{tab:quota}.
        As we can observe, compared to the population area distribution of high-quality Stack Overview questions~(see ``\% Area Quota'' column under ``Initial Seed Set''), the area distribution of final benchmark questions~(see ``\% Area Quota'' column under ``Final \sysname Benchmark'') is more balanced and less biased towards front-end, mobile, and desktop topics.
        

    \subsection{Evaluation Criteria and Evaluation Framework}
        \label{subsec:evaluation-framework-and-score-computing}

        In response to the diversified questions, \sysname evaluation framework integrates four types of model-free and automatic metrics as below.
        Domain experts choose one or multiple metric types along with their weights and concretize.
        \begin{itemize}[leftmargin=*,noitemsep]
            \item \textbf{Keywords Matching.}
            Though the responses can be in diverse forms, for a significant portion of benchmark questions, we find that the existence of some keywords strongly determines the quality of the response.
            Domain experts can write rules that match keywords and regular expressions or construct recursive logical expressions on top of keyword-matching results.
            When multiple keywords exist, each matching result can have its weight in the final score.

            \item \textbf{Blank Filling.}
            For some questions, it is challenging to measure the correctness given the response uncertainty. 
            In this case, domain experts can instruct the model to answer the question by following a given template and filling in the blanks in the template.
            The blanks can correspond to either natural language or code snippet.
            Then, similar to keywords matching, each blank can match potential keywords, regular expressions, or recursive logic expressions built upon matching results.
            This metric type tests not only the model's QA ability but also its instruction-following ability.

            \item \textbf{Unit Testing.}
            For code-intensive questions, we can follow existing benchmarks to evaluate response correctness by unit tests.
            For this type, domain experts may add more specifications in the prompt to allow for unit-test-based evaluation, such as specifications on function name, input arguments, and output format.
            Domain experts can further import the context setup and cleanup script.

            \item \textbf{Dialogue Similarity.}
            For natural-language-intensive questions, domain experts can extract and shorten the reference answers from Stack Overflow, and then use the ROUGE score~\citep{lin2004rouge} to evaluate the response similarity with reference answers.
            The ROUGE score was initially proposed and widely used in evaluating the quality of text summarization and machine translation.
            To map the ROUGE score back to our benchmark scale, we allow domain experts to tune the mapping interval and scores within the interval are then linearly mapped to our score scale.
        \end{itemize}
        The example questions and corresponding criteria are illustrated in \Cref{fig:example}.
        Detail statistics of metric type ratios, question type ratios, and prompt length are shown in \Cref{tab:benchmark-statistics}.

        \begin{wraptable}{r}{0.5\linewidth}
        \caption{\small \sysname statistics.}
        \centering
        \vspace{-1.0em}
        \begin{subfigure}[Question type.]
            {
                \resizebox{0.52\linewidth}{!}{
                    \begin{tabular}{c|c}
                        \toprule
                        Question Type & Ratio \\
                        \midrule
                        Code Completion & 30.37\% \\
                        Knowledge Question-Answering & 27.04\% \\
                        Code Debugging & 26.67\% \\
                        Config \& Environment Debugging & 15.93\% \\
                        \bottomrule
                    \end{tabular}
                }
            }
        \end{subfigure}
        \hfill
        \begin{subfigure}[Metric type.]
            {
                \resizebox{0.38\linewidth}{!}{
                    \begin{tabular}{c|c}
                        \toprule
                        Metric Type & Ratio \\
                        \midrule
                        Keywords Matching & 57.41\% \\
                        Blank Filling & 12.22\% \\
                        Unit Testing & 19.26\% \\
                        Dialogue Similarity & 11.85\% \\
                        \bottomrule
                    \end{tabular}
                }
            }
        \end{subfigure}
        \vspace{-1.0em}
        \begin{subfigure}[Prompt token length with Code Llama tokenizer.]
            {
                \resizebox{\linewidth}{!}{
                    \begin{tabular}{cccccc}
                        \toprule
                        min & 25\% quantile & median & mean & 75\% quantile & max \\
                        \midrule
                        43 & 145.75 & 223 & 338.46 & 359.50 & 5047 \\
                        \bottomrule
                    \end{tabular}
                }
            }
        \end{subfigure}
        \label{tab:benchmark-statistics}
    \end{wraptable}

        \paragraph{Score Computation.}
        We treat each question equally with one point each.
        Given 234 questions in the benchmark, the full score is 234, and we by default report the percentage score~(achieved score divided by 234) unless otherwise noted.
        The one point for each question can be further decomposed into a few scoring points within each question.
        For example, a question may contain four keywords with weights 2, 1, 1, and 1 each.
        Then, matching each keyword can contribute to 0.4, 0.2, 0.2, and 0.2 points respectively to the final score.

        \paragraph{Implementation.}
        We have implemented an automated evaluation framework with Python, publicly available at {\url{https://infi-coder.github.io/infibench}}.
        Specifically, for blank-filling evaluation, we use the longest common subsequence matching via dynamic programming to capture the filled blanks in the response.
        For unit-testing evaluation, we construct a runtime environment that supports the test execution for nine languages. 
        We plan to integrate the framework into the Hugging Face Open LLM Leaderboard~\citep{open-llm-leaderboard} to further ease the evaluation burden. 

        \paragraph{How does \sysname Evaluation Align with Human?}
        To evaluate the alignment between \sysname evaluation and human expert evaluation, we randomly sample 100 questions without replacement from the benchmark and select three strong LLMs to generate responses: GPT-4-0613, GPT-3.5-turbo, and Mistral Codestral-22b.
        For each question, we randomly choose two out of these three model responses to construct response pairs, resulting in 100 response pairs $\gR = \{(A_i, B_i): 1\le i\le 100\}$.
        For each response pair $(A, B) \in \gR$, we use \sysname, GPT-4o, and human expert to evaluate into four outcomes: $A$ is more correct than $B$~($A>B$); $B$ is more correct then $A$~($B>A$); both $A$ and $B$ are correct~($A\approx B\uparrow$); both $A$ and $B$ are incorrect~($A\approx B\downarrow$).
        \emph{Our purpose is to evaluate how \sysname evaluation aligns with humans, specifically when compared to the widely-used LLM-as-a-judge~(i.e., model-based evaluation)~\citep{zheng2023judging}.}
        The concrete grading criteria is as below:
        \begin{itemize}[leftmargin=*]
            \item \sysname gives a score between $[0,1]$ for each response in the pair. 
            If the score difference in the pair is larger than 0.2, we label the outcome to be $A>B$ or $B>A$ respectively;
            otherwise, if the maximum score among the two is larger than 0.5, we label the outcome to be $A\approx B\uparrow$; 
            otherwise, we label the outcome to be $A\approx B\downarrow$.

            \item For GPT-4o evaluation, we deploy the prompting template from LLM-Blender~\cite[Appendix E]{jiang2023llm} and trigger GPT-4o for grading the four outcomes. 
            We enhance the reliability of the comparison by switching $A$ and $B$ and prompting GPT-4o twice. 
            We record the preference only when a consistent preference exists.

            \item For human evaluation, we recruit human annotators who came up with the criteria to label the comparison preference since they are familiar with the questions and have strong expertise. 
            Annotators have no access to the evaluation results of \sysname and GPT-4o, nor which source model generates the response. 
            Annotators were instructed to directly label each pair with the four outcomes.
        \end{itemize}
        
        We defer the consensus matrices between \sysname/GPT-4o and human annotators along with more findings in \Cref{app:agreement-statistics}.
        If we only count the cases where both \sysname/GPT-4o and humans have clear preferences, the agreement rate between \sysname and humans is 85.1\%, and the agreement rate between GPT-4o and humans is 77.8\%.
        Hence, \emph{the \sysname evaluation aligns with human experts better than the GPT-4o evaluation (with >80\% confidence).} 
        We observe that
        the advantage of \sysname comes from the ability to detect deceptive answers.
        some model responses pretend to be helpful with lengthy wording and hallucinations.
        GPT-4o is more likely to be cheated than \sysname, which looks for key concepts that should exist in a helpful answer.



    \subsection{Mitigations on Memorization and Data Contamination}
        \label{subsec:data-contamination}

        \sysname is created from the publicly available Stack Overflow corpus to reflect real-world scenarios, and this corpus may already exist in the training set of some code LLMs~(e.g., DeepSeek Coder~\citep{deepseekcoder} and StarCoder 2~\citep{lozhkov2024starcoder}).
        Hence, some code LLMs may achieve a high score simply due to memorization.
        To mitigate this, we asked the domain experts to paraphrase every question as an essential step~(see \Cref{subsec:annotation}).
        Hence, copying either the highly voted answers or officially accepted answers of the original questions only achieves 65.18\%, being far from perfect and inferior to GPT-4's 70.64\%.
        Furthermore, code LLMs that use Stack Overflow data do not demonstrate significant advantages over those without.
        Hence, we deem the effect of contamination as small.

        On the other hand, we release the post IDs of the source question posts of \sysname.
        Hence, future LLM training could consider this benchmark to conduct deduplication and ablation studies on data contamination.
        Another usage of our benchmark is to evaluate retrieval-augmented~(RAG) code LLMs where perfect retrieval from Stack Overflow and moderate adaptation should solve these questions, which we leave as future work.

    \subsection{Comparison with Existing Benchmarks}
        \label{subsec:comparison-with-existing-bench}

        In \Cref{tab:benchmark-comparison}, we compare \sysname with several existing benchmarks for code LLMs.
        As reflected in the table, \sysname strongly complements existing benchmarks for code LLMs by 
        (1)~extending them beyond code generation to a wide range of real-world tasks,
        (2)~diversifying them since \sysname does not share the same source as existing ones,
        and
        (3)~increasing the differentiation as an unsaturated benchmark.
        Related benchmarks are further illustrated in \Cref{sec:related}.
        On the other hand, the benchmark is limited in size due to the high cost of correctness criteria labelling, and we are continuously expanding the benchmark.

\section{Evaluation and Leaderboard}
    \label{sec:eval-and-leaderboard}

    We systematically evaluated over 100 code LLMs spanning both proprietary and open-source worlds on \sysname.
    To the best of our knowledge, this is the most extensive evaluation for code LLMs.

    \paragraph{Evaluation Protocol.}
        \label{subsec:evaluation-protocol}
        We adopt best@10 as the main evaluation metric: 10 responses are sampled and evaluated for each question, then the best score per question is recorded and summed up.
        Throughout the evaluation, we set sampling temperature $T=0.2$ and top $p=0.9$.
        
        Furthermore, we swept sampling parameters with GPT-4 and the detailed results are in \Cref{app:gen-hyperparam}.
        In a nutshell, for maximizing the performance under best@10, the best parameters are $T=1.0$ and $p=0.9$, leading to a score of $76.15\% \pm 0.21\%$~(in comparison to $70.64\% \pm 0.82\%$ in our main setting $T=0.2, p=0.9$).
        In particular, the temperature $T$ affects much and the effect of top $p$ is minor.
        We decided to stick to the original parameters $T=0.2$ and $p=0.9$ in the main evaluation since this setting is more akin to the real-world scenario where user generates once with low temperature.

        We design two system prompts~(shown in \Cref{app:prompt-template}), one for normal questions and the other for open-ended questions with an additional sentence to encourage succinct responses.
        For generic models, we generate the prompt with ``{\small\tt \{system prompt\}\textbackslash n\{content prompt\}}'' format;
        for instruction-finetuned or chat models, we generate the prompt with their prompt templates.

        For proprietary models, we evaluate the latest models from OpenAI~(GPT-4, GPT-4o, etc), Anthropic~(Claude 3), and Mistral AI~(Mistral Small/Medium/Large) with API calling.
        When budget permits, we repeat each evaluation three times and report standard deviation. 
        For open-source models, we download models from Hugging Face and evaluate them on an 8xA100 server with bigcode-evaluation-harness~\citep{bigcode-evaluation-harness}.
        When the model size is within 30B parameters, we repeat each evaluation three times and report the standard deviation.
        All raw model responses are available at \url{https://figshare.com/articles/dataset/InfiBench_Detail_Evaluation_Data/26104864}.
        More details on the evaluation protocol are in \Cref{app:benchmark-result}.

    \vspace{-0.5em}
    \paragraph{Leaderboard.}

\begin{table}
    \caption{\small \textbf{Aggregated \sysname leaderboards (best viewed zoomed in and in color).}
     ``Size'' column records number of parameters.
     For MoE models, ``total params. / params. activated during inference'' is recorded.
     Bar colors stand for \colorbox{barblue1}{General Base}, \colorbox{barblue2}{General Finetuned}, \colorbox{barorange1}{Code Base}, and \colorbox{barorange2}{Code Finetuned} models respectively. 
     Icon ``\twemoji{locked}'' stands for proprietary models otherwise open-source. Full leaderboard in \Cref{app:benchmark-result}.}
    \label{tab:main_abstract_barplot}
    \vspace{0.3em}
    \centering
    \sbox0{\begin{subfigure}[t][\small \sysname leaderboard by model family, where best model within each model family is shown.]
    {
    \resizebox{0.48\linewidth}{!}{
        \begin{tabular}{lllcp{4.0cm}}
            \toprule
            & Family & Best Model Name & Size & \sysname Score \\
            \midrule

 1 & \twemoji{locked}GPT-4 & GPT-4-0613                & ?          & \begin{generaltunebar}{4.9448cm} \flushright \textcolor{bargreen3} {70.64\% $\pm$ 0.82\%} \end{generaltunebar} \\
 2 & DeepSeek Coder & deepSeek-coder-V2-instruct & 236B / 21B & \begin{codetunebar}{4.5843cm} \flushright \textcolor{bargreen3} {65.49\%} \end{codetunebar} \\
 3 & \twemoji{locked}Claude 3 & Claude 3 Opus             & ?          & \begin{generaltunebar}{4.472300000000001cm} \flushright \textcolor{bargreen3} {63.89\%} \end{generaltunebar} \\
 4 & Mistral Open & Codestral-22b             & 22B        & \begin{codetunebar}{4.4086cm} \flushright \textcolor{bargreen3} {62.98\% $\pm$ 0.56\%} \end{codetunebar} \\
 5 & Phind    & Phind-CodeLlama-34B-v2    & 34B        & \begin{codetunebar}{4.13cm} \flushright \textcolor{bargreen3} {59.00\%} \end{codetunebar} \\
 6 & \twemoji{locked}Mistral & mistral-large             & ?          & \begin{generaltunebar}{4.075399999999999cm} \flushright \textcolor{bargreen3} {58.22\%} \end{generaltunebar} \\
 7 & DeepSeek LLM & deepseek-llm-67b-chat     & 67B        & \begin{generaltunebar}{4.0187cm} \flushright \textcolor{bargreen3} {57.41\%} \end{generaltunebar} \\
 8 & \twemoji{locked}GPT-3.5 & GPT-3.5-turbo-0613        & ?          & \begin{generaltunebar}{3.9528999999999996cm} \flushright \textcolor{bargreen3} {56.47\% $\pm$ 1.34\%} \end{generaltunebar} \\
 9 & Qwen     & Qwen-72B                  & 72B        & \begin{generalpretrainbar}{3.8738cm} \flushright \textcolor{bargreen3} {55.34\%} \end{generalpretrainbar} \\
10 & Magicoder & Magicoder-S-CL-7B         & 7B         & \begin{codetunebar}{3.6897cm} \flushright \textcolor{bargreen3} {52.71\% $\pm$ 0.72\%} \end{codetunebar} \\
11 & WizardLM & WizardCoder-Python-34B-V1.0 & 34B        & \begin{codetunebar}{3.6813cm} \flushright \textcolor{bargreen3} {52.59\%} \end{codetunebar} \\
12 & Code Llama & CodeLlama-34b-Instruct    & 34B        & \begin{codetunebar}{3.5315000000000003cm} \flushright \textcolor{bargreen3} {50.45\%} \end{codetunebar} \\
13 & 01.AI    & Yi-34B-Chat               & 34B        & \begin{generaltunebar}{3.4706cm} \flushright \textcolor{bargreen3} {49.58\%} \end{generaltunebar} \\
14 & Zephyr   & Zephyr 7B beta            & 7B         & \begin{generaltunebar}{3.2417000000000002cm} \flushright \textcolor{bargreen3} {46.31\% $\pm$ 1.11\%} \end{generaltunebar} \\
15 & StarCoder2 & 15B-Instruct              & 15B        & \begin{codetunebar}{3.2123000000000004cm} \flushright \textcolor{bargreen3} {45.89\% $\pm$ 0.95\%} \end{codetunebar} \\
16 & DeepSeek MoE & deepseek-moe-16b-chat     & 16B / 2.8B & \begin{generaltunebar}{3.1626cm} \flushright \textcolor{bargreen3} {45.18\% $\pm$ 1.65\%} \end{generaltunebar} \\
17 & OctoPack & OctoCoder                 & 15.5B      & \begin{codetunebar}{3.1184999999999996cm} \flushright \textcolor{bargreen3} {44.55\% $\pm$ 0.79\%} \end{codetunebar} \\
18 & gemma    & gemma-7b-it               & 7B         & \begin{generaltunebar}{2.8476cm} \flushright \textcolor{bargreen3} {40.68\% $\pm$ 1.23\%} \end{generaltunebar} \\
19 & Llama 2  & Llama2-70B-Chat           & 70B        & \begin{generaltunebar}{2.7509999999999994cm} \flushright \textcolor{bargreen3} {39.30\%} \end{generaltunebar} \\
20 & InternLM & InternLM-Chat-20B         & 20B        & \begin{generaltunebar}{2.6187cm} \flushright \textcolor{bargreen3} {37.41\% $\pm$ 0.75\%} \end{generaltunebar} \\
21 & Baichuan2 & Baichuan2-13B-Chat        & 13B        & \begin{generaltunebar}{2.408cm} \end{generaltunebar} \vspace{-0.34cm} \textcolor{bargreen3} {34.40\% $\pm$ 1.34\%} \\
22 & StarCoder & StarCode+                 & 15.5B      & \begin{codepretrainbar}{2.1469cm} \end{codepretrainbar} \vspace{-0.34cm} \textcolor{bargreen3} {30.67\% $\pm$ 1.57\%} \\
23 & CodeGen2.5 & CodeGen2.5-7B-Instruct    & 7B         & \begin{codetunebar}{2.0699cm} \end{codetunebar} \vspace{-0.34cm} \textcolor{bargreen3} {29.57\% $\pm$ 1.53\%} \\
24 & ChatGLM  & ChatGLM3-6B               & 6B         & \begin{generaltunebar}{1.9761000000000002cm} \end{generaltunebar} \vspace{-0.34cm} \textcolor{bargreen3} {28.23\% $\pm$ 0.58\%} \\
25 & \twemoji{locked}davinci & davinci-002               & ?          & \begin{generalpretrainbar}{1.4875cm} \end{generalpretrainbar} \vspace{-0.34cm} \textcolor{bargreen3} {21.25\% $\pm$ 1.17\%} \\
26 & Phi      & Phi1.5                    & 1.5B       & \begin{generalpretrainbar}{1.4391999999999998cm} \end{generalpretrainbar} \vspace{-0.34cm} \textcolor{bargreen3} {20.56\% $\pm$ 0.09\%} \\
27 & CodeGeeX & CodeGeeX2-6B              & 6B         & \begin{codepretrainbar}{1.3916cm} \end{codepretrainbar} \vspace{-0.34cm} \textcolor{bargreen3} {19.88\% $\pm$ 0.36\%} \\
28 & CodeGen2 & CodeGen2-16B              & 16B        & \begin{codepretrainbar}{1.1879cm} \end{codepretrainbar} \vspace{-0.34cm} \textcolor{bargreen3} {16.97\% $\pm$ 1.15\%} \\
29 & IEITYuan & Yuan2-51B-hf              & 51B        & \begin{generalpretrainbar}{1.0675cm} \end{generalpretrainbar} \vspace{-0.34cm} \textcolor{bargreen3} {15.25\%} \\
30 & CodeGen  & CodeGen-16B-multi         & 16B        & \begin{codepretrainbar}{0.9533999999999999cm} \end{codepretrainbar} \vspace{-0.34cm} \textcolor{bargreen3} {13.62\% $\pm$ 1.18\%} \\
\hline
\\
  \multicolumn{2}{c}{\multirow{3}{*}{Human}}    & 10 Highest-Voted Answer Posts              &           & \begin{defaultbar}{4.562600000000001cm} \flushright \textcolor{bargreen3} {65.18\%} \end{defaultbar} \\
 &     & Highest-Voted Answer Post      &           & \begin{defaultbar}{3.9396000000000004cm} \flushright \textcolor{bargreen3} {56.28\%} \end{defaultbar} \\
 &     & Officially-Accepted Answer Post           &           & \begin{defaultbar}{3.7030000000000003cm} \flushright \textcolor{bargreen3} {52.90\%} \end{defaultbar} \\

            \bottomrule
        \end{tabular}
    }
    }
    \end{subfigure}}
    %
    %
    %
    \usebox0\begin{minipage}[b][\ht0][s]{\wd0}
    \vspace{0.5em}
    \begin{subfigure}[b][\small \sysname leaderboard by model type, where top five model within each type is shown.]
    {
    \resizebox{\linewidth}{!}{
        \begin{tabular}{lllcp{4.6cm}}
            \toprule
            Type & Rank & Model Family / Model Name & Size & \sysname Score \\
            \midrule

\multirow{5}{*}{\shortstack{Pro-\\prie-\\tary\\Model}} &  1 & \twemoji{locked}GPT-4/GPT-4-0613   & ?          & \begin{generaltunebar}{4.9448cm} \flushright \textcolor{bargreen3} {70.64\% $\pm$ 0.82\%} \end{generaltunebar} \\
 &  2 & \twemoji{locked}GPT-4/GPT-4-turbo-1106 & ?          & \begin{generaltunebar}{4.7894cm} \flushright \textcolor{bargreen3} {68.42\% $\pm$ 0.38\%} \end{generaltunebar} \\
 &  3 & \twemoji{locked}GPT-4/GPT-4o-2024-05-13 & ?          & \begin{generaltunebar}{4.6333cm} \flushright \textcolor{bargreen3} {66.19\%} \end{generaltunebar} \\
 &  4 & \twemoji{locked}Claude 3/Claude 3 Opus & ?          & \begin{generaltunebar}{4.472300000000001cm} \flushright \textcolor{bargreen3} {63.89\%} \end{generaltunebar} \\
 &  5 & \twemoji{locked}Mistral/mistral-large & ?          & \begin{generaltunebar}{4.075399999999999cm} \flushright \textcolor{bargreen3} {58.22\%} \end{generaltunebar} \\
\hline
\multirow{5}{*}{\shortstack{Code\\Fine-\\tuned\\Model}}
 &  1 & DeepSeek Coder/deepSeek-coder-V2-instruct & 236B / 21B & \begin{codetunebar}{4.5843cm} \flushright \textcolor{bargreen3} {65.49\%} \end{codetunebar} \\
 &  2 & Mistral Open/Codestral-22b         & 22B        & \begin{codetunebar}{4.4086cm} \flushright \textcolor{bargreen3} {62.98\% $\pm$ 0.56\%} \end{codetunebar} \\
 &  3 & DeepSeek Coder/deepseek-coder-33b-instruct & 33B        & \begin{codetunebar}{4.4072000000000005cm} \flushright \textcolor{bargreen3} {62.96\%} \end{codetunebar} \\
 &  4 & Phind/Phind-CodeLlama-34B-v2       & 34B        & \begin{codetunebar}{4.13cm} \flushright \textcolor{bargreen3} {59.00\%} \end{codetunebar} \\
 & 5 & Phind/Phind-CodeLlama-34B-v1       & 34B        & \begin{codetunebar}{4.092899999999999cm} \flushright \textcolor{bargreen3} {58.47\%} \end{codetunebar} \\

\hline
\multirow{5}{*}{\shortstack{Code\\Base\\Model}} 
 & 1 & Code Llama/CodeLlama-34b           & 34B        & \begin{codepretrainbar}{3.3152cm} \flushright \textcolor{bargreen3} {47.36\%} \end{codepretrainbar} \\
 & 2 & Code Llama/CodeLlama-34b-Python    & 34B        & \begin{codepretrainbar}{3.0191000000000003cm} \flushright \textcolor{bargreen3} {43.13\%} \end{codepretrainbar} \\
 & 3 & StarCoder2/15B                     & 15B        & \begin{codepretrainbar}{2.9764000000000004cm} \flushright \textcolor{bargreen3} {42.52\% $\pm$ 1.24\%} \end{codepretrainbar} \\
 & 4 & Code Llama/CodeLlama-13b           & 13B        & \begin{codepretrainbar}{2.9162cm} \flushright \textcolor{bargreen3} {41.66\% $\pm$ 0.84\%} \end{codepretrainbar} \\
 & 5 & Code Llama/CodeLlama-13b-Python    & 13B        & \begin{codepretrainbar}{2.8917cm} \flushright \textcolor{bargreen3} {41.31\% $\pm$ 0.90\%} \end{codepretrainbar} \\
\hline
\multirow{5}{*}{\shortstack{General\\Fine-\\tuned\\Model}}
 & 1 & DeepSeek LLM/deepseek-llm-67b-chat & 67B        & \begin{generaltunebar}{4.0187cm} \flushright \textcolor{bargreen3} {57.41\%} \end{generaltunebar} \\
 & 2 & Mistral Open/mixtral-8x7B-Instruct & 46.7B / 12.9B & \begin{generaltunebar}{3.8884999999999996cm} \flushright \textcolor{bargreen3} {55.55\%} \end{generaltunebar} \\
 & 3 & Qwen/Qwen-72B-Chat                 & 72B        & \begin{generaltunebar}{3.7078999999999995cm} \flushright \textcolor{bargreen3} {52.97\%} \end{generaltunebar} \\
 & 4 & 01.AI/Yi-34B-Chat                  & 34B        & \begin{generaltunebar}{3.4706cm} \flushright \textcolor{bargreen3} {49.58\%} \end{generaltunebar} \\
 & 5 & Zephyr/Zephyr 7B beta              & 7B         & \begin{generaltunebar}{3.2417000000000002cm} \flushright \textcolor{bargreen3} {46.31\% $\pm$ 1.11\%} \end{generaltunebar} \\
\hline
\multirow{5}{*}{\shortstack{General\\Base\\Model}}
 & 1 & Qwen/Qwen-72B                      & 72B        & \begin{generalpretrainbar}{3.8738cm} \flushright \textcolor{bargreen3} {55.34\%} \end{generalpretrainbar} \\
 & 2 & Qwen/Qwen-14B                      & 14B        & \begin{generalpretrainbar}{3.0583cm} \flushright \textcolor{bargreen3} {43.69\% $\pm$ 1.09\%} \end{generalpretrainbar} \\
 & 3 & DeepSeek LLM/deepseek-llm-67b-base & 67B        & \begin{generalpretrainbar}{2.7908999999999997cm} \flushright \textcolor{bargreen3} {39.87\%} \end{generalpretrainbar} \\
 & 4 & Llama 2/Llama2-70B                 & 70B        & \begin{generalpretrainbar}{2.6382999999999996cm} \flushright \textcolor{bargreen3} {37.69\%} \end{generalpretrainbar} \\
 & 5 & Qwen/Qwen-7B                       & 7B         & \begin{generalpretrainbar}{2.2183cm} \end{generalpretrainbar} \vspace{-0.34cm} \textcolor{bargreen3} {31.69\% $\pm$ 0.29\%} \\

            \bottomrule
        \end{tabular}
    }
    }
    \end{subfigure}
    %
    %
    %
    \begin{subfigure}[\small \sysname leaderboard by model size, where best model within the threshold is shown.]
    {
    \resizebox{\linewidth}{!}{
        \begin{tabular}{llcp{5.0cm}}
            \toprule
            Size Threshold & Model Family / Model Name & Size & \sysname Score \\
            \midrule

 $\infty$ & \twemoji{locked}GPT-4/GPT-4-0613   & ?          & \begin{generaltunebar}{4.9448cm} \flushright \textcolor{bargreen3} {70.64\% $\pm$ 0.82\%} \end{generaltunebar} \\
 <100B & Mistral Open/Codestral-22b         & 22B        & \begin{codetunebar}{4.4086cm} \flushright \textcolor{bargreen3} {62.98\% $\pm$ 0.56\%} \end{codetunebar} \\
 <20B & DeepSeek Coder/deepseek-coder-6.7b-instruct & 6.7B       & \begin{codetunebar}{3.7275cm} \flushright \textcolor{bargreen3} {53.25\% $\pm$ 0.40\%} \end{codetunebar} \\
 <5B & DeepSeek Coder/deepseek-coder-1.3b-instruct & 1.3B       & \begin{codetunebar}{2.8924000000000003cm} \flushright \textcolor{bargreen3} {41.32\% $\pm$ 1.12\%} \end{codetunebar} \\

            \bottomrule
        \end{tabular}
    }
    }
    \end{subfigure}
    \end{minipage}

\end{table}

        In \Cref{tab:main_abstract_barplot}, we present aggregated \sysname leaderboards by model family, model type, and model size.
        The full leaderboard is deferred to \Cref{app:benchmark-result} due to space limit.
        The table includes scores from using the original Stack Overflow answer posts as reference.
        Results are also presented as a scatter plot in \Cref{fig:full-result},
        where normal models are shown as scatters with error bars, MoE models are shown as horizontal segments with error ranges connecting the activated parameters during inference and total parameters, and strong proprietary models are shown as horizontal lines.
        
        In both tables and the figure, we classify LLMs by general/code and base/finetuned.
        The general LLMs are claimed to have strong capabilities beyond code, e.g., in various natural language tasks, while the code LLMs are exclusively optimized for the code domain.
        The base LLMs only went through the pretraining phase, while the finetuned LLMs are claimed to have instruction-following capabilities or are finetuned on instruction or human preference datasets.

\section{Analysis and Discussion}
    \label{sec:analysis-discussion}
        \label{subsec:results-and-analysis}

        \begin{figure*}[!t]
            \centering
            \includegraphics[width=\linewidth]{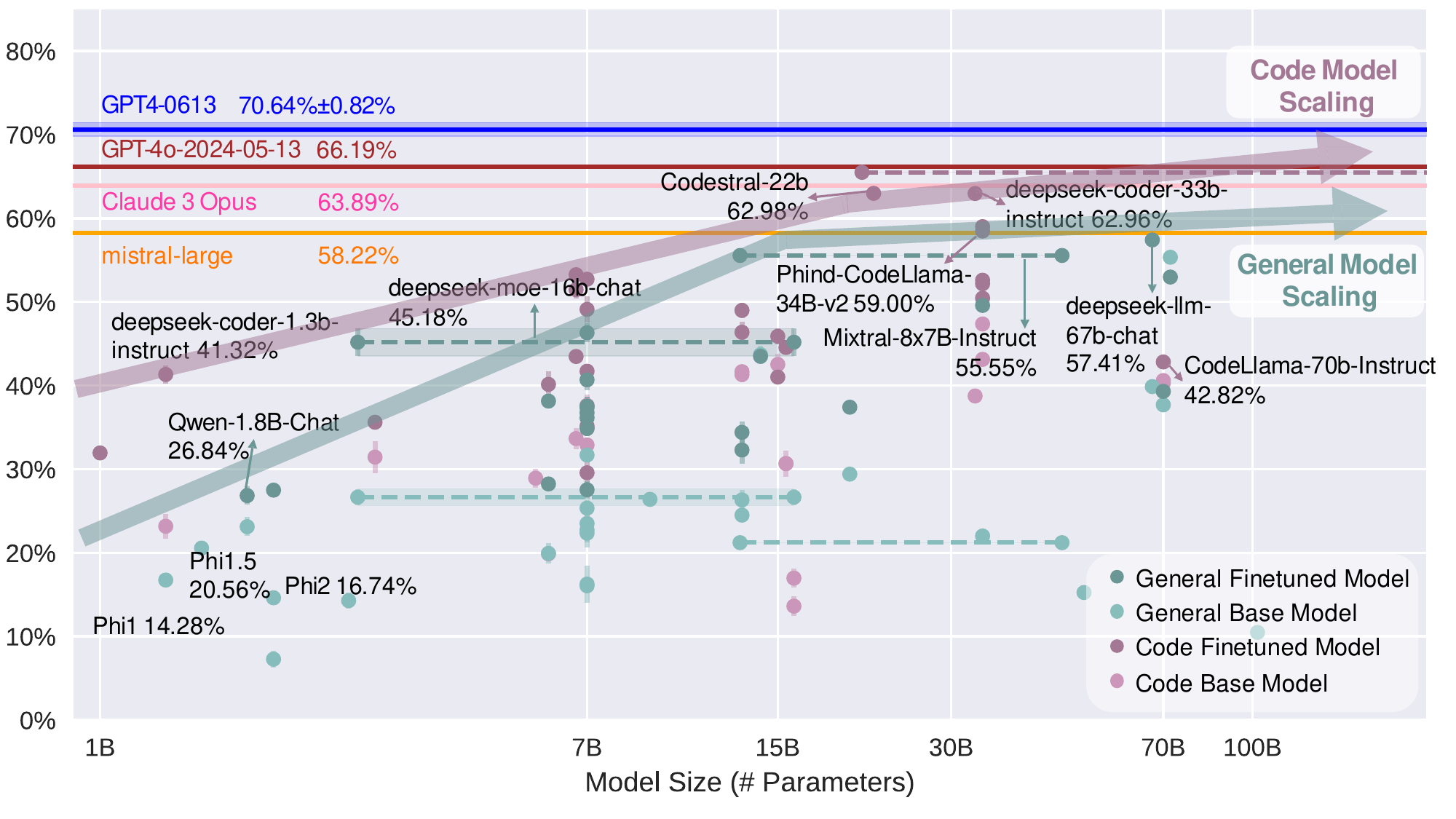}
            \caption{\small \textbf{Scatter plot for all evaluated LLMs on \sysname}.
            $x$-axis is the model size in terms of number of parameters and $y$-axis is \sysname score.
            Projected empirical scaling laws for both general and code models are drawn. 
            Detail discussion in \Cref{subsec:results-and-analysis}.
            }
            \label{fig:full-result}
            \vspace{-1em}
        \end{figure*}

        \textbf{The best model so far, GPT-4, is still far from perfect, and open-source models are competitive but still far from GPT-4.}
        GPT-4 achieves the highest score 70.64\%~(interestingly, achieved by GPT-4-0613 instead of the more recent GPT-4o), then Claude 3 Opus with a score 63.89\%, and then Codestral-22b~\citep{codestral} with a score 62.98\% and deepseek-coder-33b-instruct~\citep{deepseekcoder} with a score 62.96\%.
        The result implies that:
        (1)~Noting that the full score is 100\%, even the powerful GPT-4 is still far from perfect, which is in contrast to its $\approx$90\% HumanEval score.
        We inspect the score breakdown.
        For the two most frequent metric types, keywords matching and unit testing, GPT-4 achieves similar scores 66.61\% and 76.00\% respectively.
        For blank filling, the score is relatively lower at 58.08\%.
        These scores imply that GPT-4 may still lack generic ability in answering diversified real-world questions related to code.
        When instructed to follow a given template to answer~(blank filling), due to the more strict requirement and narrower solution space, its lack of capability is more pronounced.
        (2)~There is still a visible gap between open-source models and GPT-4.
        The gap between the most powerful open-source model, Codestral-22b, and GPT-4 is roughly 8 points.
        On the other hand, noticing that GPT-3.5-turbo achieves 56.47\%, the open-source model, Codestral-22b, is now reliably better than GPT-3.5-turbo with merely 22B parameters which is promising.
        
        \textbf{Among open-source models, different models have various performances.}
        \Cref{fig:full-result} systematically visualizes the performance of different open-source models at diverse scales.
        Although there is a general tendency that larger models achieve higher scores, the scores among different models at a similar scale differ largely.
        For example, on scale 7B, the best-performing model is at around 55\%, pretty close to GPT-3.5, while the low-performing model stays at around 15\%.
        Moreover, deepseek-coder-1.3b-instruct achieves 41.32\% at 1.3B and surpasses a few models at scale 70B or 100B.
        Hence, though scaling matters, the training techniques and training data are equally important or even more, helping to reduce the required scale for achieving a certain score by more than 10$\times$.
        

        \textbf{Hard problems generalize their difficulties.}
        We rate the benchmark problem difficulty with five levels by how well GPT-4 and GPT-3.5-turbo answer them, as detailed in \Cref{app:difficulty-grouping}.
        Example questions from each level are shown in \Cref{app:examples}.
        We present the detail result table including the sub-score for each difficulty level in \Cref{app:benchmark-result}.
        Interestingly, the trend is \emph{highly consistent that sub-scores decrease along with the increase of problem level}.
        Specifically,
        hard problems for the most powerful model yet, GPT-4, are also generally hard for open-source models.
        These hard problems usually correspond to code generation with long and domain-specific context or challenging blank-filling questions since blank-filling is a specific task that rarely appears in training data before.

        \textbf{Instruction finetuning is important for QA.}
        Among models of similar scales and the same family, we find that the best-performing ones almost always include an instruction-finetuning phase, such as deepseek-llm-67b-chat, deepseek-coder-33b-instruct, CodeLlama-34B-Instruct, and Qwen-18B-Chat.
        In contrast, the pretraining models, such as davinci-002 and phi models, usually perform poorly despite their strong performances in code generation benchmarks.
        %
        Instruction-finetuning is also critical for other code domain tasks such as code generation.
        As shown in \Cref{subapp:correlation-study} where we plot model scores in QA~(measured by \sysname) and code generation~(measured by HumanEval pass@1), instruction-tuning generally improves both QA and code generation, but the improvement is usually more significantly on code generation but more moderately on QA.
        As a result, we suggest generalizing the instruction-finetuning data beyond simple coding problems to improve code LLMs.
        Indeed, our preliminary experiments show that, after fine-tuning with the decontaminated and sanitized Stack Overflow data, we improved \sysname scores for Codellama-13b-Instruct from 46.37\% to 60.74\% and for mixtral-8x7B-Instruct from 55.55\% to 62.61\%.

        \textbf{Some models may focus too much on code generation, especially the small ones.}
        As detailed in \Cref{subapp:correlation-study}, we observe that for large models~($>$30B) and top entries, \sysname and HumanEval pass@1 scores coincide well.
        However, for smaller models, the score tendencies start to diverge, where some models are relatively stronger on \sysname~(Mixtral-8x7B-Instruct) and more are relatively stronger on HumanEval~(Phi1, Phi2, gemma-7b, ...).
        This phenomenon implies that a few models may be optimized too heavily on code generation benchmarks while ignoring the performance in generic code scenarios as represented by \sysname, which in turn highlights the significance of free-form QA benchmarks like \sysname in detecting capability imbalance in code LLMs.
        
        \begin{wraptable}{r}{0.6\linewidth}
            \caption{\small Evaluation on eight models from the Code Llama~\citep{roziere2023code} family showcases intense Python finetuning may hurt free-form QA ability, despite achieving higher HumanEval scores.}
            \resizebox{\linewidth}{!}{
                \begin{tabular}{c|c|c|cc}
                    \toprule
                    & Benchmark & Base & {Python} & {Instruct} \\
                    \midrule
                    \multirow{2}{*}{7B} & HumanEval & $33.5\%$ & $38.4\%$ (\up{$+4.9\%$}) & $34.8\%$ (\up{$+1.3\%$})  \\ 
                    & \sysname & \witherr{$37.62\%$}{$1.28\%$} & \witherr{$32.89\%$}{$0.45\%$} (\down{$-4.73\%$}) & \witherr{$35.15\%$}{$1.28\%$} (\down{$-2.47\%$}) \\
                    \hline
    
                    \multirow{2}{*}{13B} & HumanEval & $36.0\%$ & $43.3\%$ (\up{$+7.3\%$}) & $42.7\%$ (\up{$+6.7\%$}) \\
                    & \sysname &  \witherr{$41.66\%$}{$0.84\%$} & \witherr{$41.31\%$}{$0.90\%$} (\down{$-0.35\%$}) &  \witherr{$46.37\%$}{$1.26\%$} (\up{$+4.71\%$}) \\
                    \hline
                    
                    \multirow{2}{*}{34B} & HumanEval & $48.8\%$ & $53.7\%$ (\up{$+4.9\%$}) & $41.5\%$ (\down{$-7.3\%$}) \\
                    & \sysname & $47.36\%$ & $43.13\%$ (\down{$-4.23\%$}) & $50.45\%$ (\up{$+3.09\%$}) \\
                    \hline

                    \multirow{2}{*}{70B} & HumanEval & $53.0\%$ & $57.3\%$ (\up{$+4.3\%$}) & $67.8\%$ (\up{$+14.8\%$}) \\
                    & \sysname & $40.60\%$ & $40.29\%$ (\down{$-0.31\%$}) & $42.82\%$ (\up{$+2.22\%$}) \\
                    
                    \bottomrule
                \end{tabular}
            }
            
            \label{tab:codellama-case-study}
        \end{wraptable}

        \textbf{Code LLama models have unique characteristics.}
        We evaluated all Code Llama models~\citep{roziere2023code}.
        As shown in \Cref{tab:codellama-case-study}, we found finetuning on Python data improves on HumanEval but hurts \sysname scores, while instruction finetuning usually improves \sysname scores but may hurt HumanEval.
        As a side product, we found CodeLlama-70B may be overly safeguarded and denies answering some safe questions in \sysname.
        More model-specific findings are presented in \Cref{app:additional-discussion}.

        \textbf{Code models and general models may exhibit different scaling laws, and open-source models scale well only within 40B yet.}
        In \Cref{fig:full-result}, we use the top-performing code and general models at each scale respectively to regress and extrapolate model performance at larger scales.
        As shown, code models tend to have higher capabilities compared to general models of the same scale, though the gap shrinks for larger models.
        Hence, when the compute budget is heavily limited, training exclusively in the code domain could be more efficient for building strong code LLMs.

        In \Cref{fig:full-result}, both predicting curves are split into two segments, steep in the first segment and much flat in the second.
        Following the first segment, open-source models catch up with GPT-4 at around 50B scale.
        However, following the second segment, they may need to be at >300B scale to catch up.
        The finding contradicts the common scaling law~\citep{kaplan2020scaling,muennighoff2024scaling,bi2024deepseek} where a strong linear relationship between model scale and capability exists.
        The contradiction implies that very large open-source models~(>40B) may fail to achieve the expected performance at their scales, or there is some non-trivial barrier when scaling the model beyond 40B, or the scaling law may change at such a large scale.
        We leave further investigation as the future work.
        Notably, after the release of \sysname, Deepseek-coder-v2~\citep{zhu2024deepseek} was released as the largest code LLM to our knowledge in an MoE architecture with 236B total and 21B active parameters. On \sysname, Deepseek-coder-v2 achieves 65.49\%, setting the new baseline for open-source LLMs but still being inferior to GPT-4.
        More importantly, the score is within the predicted range of our empirical scaling law. 
        

        We defer dataset card and data accessibility details, discussion on limitations and societal impact, full leaderboard, additional findings, ablation studies, and data examples in appendices.





    









\vspace{-1em}
\section{Related Work}
\vspace{-1em}
    \label{sec:related}

    Large language models~\citep{vaswani2017attention,devlin2018bert,brown2020language} are transforming people's lives.
    In the coding domain, LLMs~\citep{chen2021evaluating,li2022competition} are shown to be capable of completing a wide range of tasks such as code generation, debugging, and question-answering.
    Recently, code LLMs are booming. 
    New models, including both proprietary~\citep{github2023copilot,OpenAI2023GPT4TR} and open-source ones~\citep{open-llm-leaderboard,nijkamp2023codegen,touvron2023llama1,touvron2023llama,li2023starcoder,luo2023wizardcoder,roziere2023code,zhu2024deepseek}, emerge almost every month.

    Benchmarks for code LLMs are developing, though at a relatively slower pace.
    Common benchmarks, e.g., APPS~\citep{hendrycks2021measuring}, MBPP~\citep{austin2021program}, and HumanEval~\citep{chen2021evaluating}, focus on code generation and unit-test-based evaluation.
    Some efforts augment these benchmarks by language translation~(e.g., Multilingual HumanEval~\citep{athiwaratkun2023multilingual}, HumanEval-X~\citep{zheng2023codegeex}), test augmentation~(e.g., HumanEval+~\citep{liu2023is}), task generalization~(e.g., HumanEvalPack~\citep{muennighoff2023octopack}), and human rewriting~(e.g., LBPP~\citep{matton2024leakage}).
    To systematically evaluate real-world problem solving, recently, SWE-bench~\citep{jimenez2024swe}, its filtered version SWE-bench Verified~\citep{openai2024sweverified}, and RepoBench~\citep{liu2023repobench} are proposed but they still primarily focus on code generation.
    Some general-purpose benchmarks, e.g., Arena-Hard~\citep{li2024crowdsourced}, contain code-related questions, but rely on LLM to judge and do not provide domain-specific scores.
    CodeXGLUE~\citep{lu12021codexglue} considers multiple coding capabilities beyond code generation, but replies on existing data sources.
    In contrast to these benchmarks, \sysname benchmark is built for evaluating free-form question-answering ability in the code domain beyond code generation in an automated and model-independent way.

\vspace{-1em}
\section{Conclusion}
\vspace{-1em}

We proposed \sysname, a systematic benchmark for evaluating the question-answering ability of code LLMs in real-world scenarios, to facilitate development and scientific evaluation of LLMs.
\sysname comprises 234 high-quality questions from Stack Overflow and supports automatic model-free evaluation.
A comprehensive evaluation of over 100 code LLMs reveals several findings and takeaways.
The benchmark is publicly available and continuously expanding.




\section*{Acknowledgement}
We thank ByteDance Inc. for the support on computing resources, anonymous reviewers for their constructive feedback, and Kaixin Li~(National University of Singapore) for contributing the Docker image after the initial release of \sysname. 
This work was partially supported by National Natural Science Foundation of China under Grant No. 62161146003, and the Tencent Foundation/XPLORER PRIZE. 
Tao Xie is also affiliated with the School of Computer Science, Peking University, China. The corresponding authors are Linyi Li and Tao Xie.
    
\bibliographystyle{plainnat}
\bibliography{root}

\section*{Checklist}


\begin{enumerate}

\item For all authors...
\begin{enumerate}
  \item Do the main claims made in the abstract and introduction accurately reflect the paper's contributions and scope?
    \answerYes{}
  \item Did you describe the limitations of your work?
    \answerYes{Limitations are discussed throughout \Cref{sec:creation} and specifically in \Cref{app:limitation}.}
  \item Did you discuss any potential negative societal impacts of your work?
    \answerYes{Societal impacts are discussed in \Cref{app:limitation}.}
  \item Have you read the ethics review guidelines and ensured that your paper conforms to them?
    \answerYes{}
\end{enumerate}

\item If you are including theoretical results...
\begin{enumerate}
  \item Did you state the full set of assumptions of all theoretical results?
    \answerNA{}
	\item Did you include complete proofs of all theoretical results?
    \answerNA{}
\end{enumerate}

\item If you ran experiments (e.g. for benchmarks)...
\begin{enumerate}
  \item Did you include the code, data, and instructions needed to reproduce the main experimental results (either in the supplemental material or as a URL)?
    \answerYes{All code and data are publicly available at \url{https://infi-coder.github.io/infibench} along with instructions needed to reproduce. The accessibility information is also available in detail in \Cref{app:dataset-card}.}
  \item Did you specify all the training details (e.g., data splits, hyperparameters, how they were chosen)?
    \answerNA{This work does not involve model training. The inference hyperparameters are listed in \Cref{sec:eval-and-leaderboard} and ablation studies are presented in \Cref{app:gen-hyperparam}.}
	\item Did you report error bars (e.g., with respect to the random seed after running experiments multiple times)?
    \answerYes{All experiments are repeated three times whenever budget and computing resource permit. Error bars and standard deviations are reported.}
	\item Did you include the total amount of compute and the type of resources used (e.g., type of GPUs, internal cluster, or cloud provider)?
    \answerYes{The information is included in \Cref{sec:eval-and-leaderboard}.}
\end{enumerate}

\item If you are using existing assets (e.g., code, data, models) or curating/releasing new assets...
\begin{enumerate}
  \item If your work uses existing assets, did you cite the creators?
    \answerYes{}
  \item Did you mention the license of the assets?
    \answerYes{The main assets are from Stack Overflow which is open source under CC BY-SA 4.0 license. We inherit this license to release.}
  \item Did you include any new assets either in the supplemental material or as a URL?
    \answerYes{New assets~(the benchmark and evaluation tool) is accessible through \url{https://infi-coder.github.io/infibench}.}
  \item Did you discuss whether and how consent was obtained from people whose data you're using/curating?
    \answerYes{We release the new asset inheriting the CC BY-SA 4.0 license as described in \Cref{sec:intro} and \Cref{app:dataset-card}.}
  \item Did you discuss whether the data you are using/curating contains personally identifiable information or offensive content?
    \answerYes{Domain experts are required to remove such information by paraphrasing when constructing the benchmark.}
\end{enumerate}

\item If you used crowdsourcing or conducted research with human subjects...
\begin{enumerate}
  \item Did you include the full text of instructions given to participants and screenshots, if applicable?
    \answerNA{}
  \item Did you describe any potential participant risks, with links to Institutional Review Board (IRB) approvals, if applicable?
    \answerNA{}
  \item Did you include the estimated hourly wage paid to participants and the total amount spent on participant compensation?
    \answerNA{}
\end{enumerate}

\end{enumerate}

\newpage

\appendix

\appendix
\onecolumn

\section*{Appendices}

In appendices, we present
dataset card and data accessibility details in \Cref{app:dataset-card}, discussion on limitations and societal impact in \Cref{app:limitation}, agreement statistics between \sysname/GPT-4o and human in \Cref{app:agreement-statistics}, question grouping by difficulty in \Cref{app:difficulty-grouping}, full leaderboard in \Cref{app:benchmark-result}, additional findings in \Cref{app:additional-discussion}, study of sampling hyperparameters in \Cref{app:gen-hyperparam}, prompts in \Cref{app:prompt-template}, and benchmark data examples in \Cref{app:examples}.

\section{Dataset Card and Accessibility Details}
    \label{app:dataset-card}

    \begin{dsbox}
        \begin{itemize}[leftmargin=*]

            \item \textbf{Name}: InfiBench

            \item \textbf{Description}: Evaluation Dataset for the Question-Answering Capabilities of Code Large Language Models

            \item \textbf{URL}: \url{https://infi-coder.github.io/infibench} (all resources) / \url{https://huggingface.co/datasets/llylly001/InfiBench} (data part)

            \item \textbf{Version}: 2.1

            \item \textbf{License}: Creative Commons Attribution Share Alike 4.0

            \item \textbf{Citation}:

            \begin{verbatim}         
@misc{infibench,
    title={InfiBench: Evaluating the Question-Answering Capabilities 
    of Code Large Language Models},
    howpublished = "\url{https://infi-coder.github.io/infibench}",
    author={InfiBench},
    year={2024}
}\end{verbatim}

            \item \textbf{DOI}: \texttt{doi:10.57967/hf/2474}

            \item \textbf{Responsible AI --- Data Collection}: \\
Data source is downloaded from the publicly available StackExchange archive (\url{https://archive.org/download/stackexchange}, \url{https://ia904700.us.archive.org/view_archive.php?archive=/6/items/stackexchange/stackoverflow.com-Posts.7z}). Especially, we use the preprocessed version from \url{https://huggingface.co/datasets/mikex86/stackoverflow-posts} where all posts are formatted in Markdown text.

We choose to keep only the questions with at least three positively voted answers and an officially accepted answer, which turn out to be 1,090,238 questions. For these one million questions, we choose to keep frequently viewed and relatively new questions.

Utilizing the mandatory question tags of these questions, we then manually construct a tag tree that covers the 200 most frequent tags, enabling us to identify the top programming languages and areas for 14,330105 out of these 17,402 questions. We exclude 6 programming languages that either describe data or are domain-specific: JSON, regex, Markdown, YAML, CSV, and SQL. As a result, we compile 13,854 questions that serve as the initial seed set.

We randomly sample from the initial seed set. Then we recruited five domain experts inside our company to create the benchmark from the sampled initial seed set, each in charge of one area. The annotation process is composed of three steps: (1) Question Selection and Type Annotation; (2) Prompt Paraphrasing. (3) Correctness Criterion Annotation.

            \item \textbf{Responsible AI --- Data Biases}: \\
The data essentially serves as an evaluation benchmark. We foresee data biases in the following aspects:

(1) Non-standard evaluation. Alongside the data is a comprehensive benchmark of existing code LLMs. The benchmark scores are evaluated under a specific set of hyperparameters (e.g, temperature 0.2, top probability 0.9, best@10 at question level). Data usage under different evaluation conditions may result in misleading comparison results and conclusions.

(2) Usage misinterpretation. The benchmark focuses on evaluating the response correctness of code LLMs for a set of real-world developers' questions. Our evaluation standard does not specifically take other aspects (naturalness, conciseness, fairness, politeness, etc) into consideration. Hence, this is risk of overinterpreting the evaluation results. When evaluating a code LLM, we recommend combining this benchmark score with other evaluations to be a more comprehensive evaluation.

(3) Potential data contamination. Though we have made our efforts to reduce the impact of data contamination, future code LLMs may train or fine-tune on this benchmark dataset to improve the score on InfiBench. This could be challenging to prevent as a cost of being fully public. On the other hand, as responsible LLM developers, we hope future practitioners would report how they use the benchmark data if beyond the original scope (for evaluation use).

            \item \textbf{Responsible AI --- Personal Sensitive Information}:
During the data construction process, our domain experts paraphrased the question prompts to remove personal and sensitive information (PII) and a cross validation stage was introduced to further ensure the PII removal.
            
        \end{itemize}
    \end{dsbox}

    \paragraph{Croissant Dataset Description:} \url{https://huggingface.co/datasets/llylly001/InfiBench/blob/main/croissant-infibench.json}.
    Note that the Croissant format is mainly designed for machine learning dataset description.
    However, \sysname is more than a dataset; it is an evaluation benchmark including response evaluation standards, tools, and an accompanying leaderboard.
    Hence, the Croissant script records only the CSV file and covers question prompts and evaluation standards; whereas the open-source evaluation tool and leaderboard are not recorded which can be separately downloaded from \url{https://infi-coder.github.io/infibench}.

    \paragraph{Data Accessibility.}
    As briefly mentioned in the main text, all materials are made publicly available and accessible at the website: \url{https://infi-coder.github.io/infibench} without personal request.
    The materials include three parts: 
    (1)~Benchmark questions and evaluation metrics --- this part is additionally uploaded to Hugging Face~(URL and DOI are in the above dataset card).
    (2)~Automatic evaluation tool --- this part is uploaded and maintained in a dedicated GitHub repo \texttt{https://github.com/infi-coder/infibench-evaluator}.
    In addition, we uploaded our extension of \texttt{bigcode-evaluation-harness}~\citep{bigcode-evaluation-harness}, namely \texttt{infibench-evaluation-harness} to a dedicated GitHub repo \url{https://github.com/infi-coder/infibench-evaluation-harness}. The extension includes the inference code on \sysname for all evaluated LLMs.
    (3)~Evaluation raw data and leaderboard --- the leaderboard is displayed on the website \url{https://infi-coder.github.io/infibench} and the raw model responses are stored in the website repo \url{https://github.com/infi-coder/infibench}.
    All materials are under the Creative Commons Attribution Share Alike 4.0 license.
    In the above dataset card and \Cref{app:limitation}, we anticipate potential inappropriate usage of the benchmark and we encourage the practitioners to document their usage of the benchmark if beyond model evaluation.
    In the future, we will continue the maintenance and expansion of the benchmark.
    Furthermore, we are developing an adaptor for automatic evaluation on Hugging Face so that \sysname can be integrated into the Hugging Face Open LLM Leaderboard~\citep{open-llm-leaderboard} to further ease the evaluation burden. 

\section{Limitations, Societal Impacts, and Future Work}
    \label{app:limitation}

    In this appendix, we expand our discussion of limitations, potential societal impacts, and future work.

    \paragraph{Evaluation Metric.}
    In \sysname, the expert-annotated evaluation metric is designed to mainly focus on response correctness, more specifically, whether the response contains key information that solves the given question.
    Concretely, the metric may evaluate whether the response passes a given set of unit tests, whether it suggests the right API or concept, whether it follows the instruction to provide relevant information, etc.
    Hence, the score comes with two limitations:
    (1)~The score is subjective since the metric is annotated by human experts without an explicit and universal standard.
    Note that we did not aim to provide an objective metric since the developers' views of response correctness intrinsically vary and diverge for these diverse questions.
    On the other hand, we introduce a cross-validation and calibration stage to improve the metric representativeness of most developers' standards.
    We leave it as a future work to further quantitatively measure and improve the metric representativeness.
    (2)~The score focuses mainly on correctness.
    Several other aspects define a model's usability, such as language naturalness~(including conciseness, politeness, etc), trustworthiness~(refusal of risky questions, fairness, unbiasedness, privacy, etc), and system-level metrics~(latency, throughput, parallelism-friendliness, etc).
    Model evaluators and practitioners may keep in mind that \sysname score is not a comprehensive usability measurement of code LLMs, and we strongly encourage them to combine \sysname score with benchmarks on these other aspects~(c.f. \citep{bhatt2024cyberseceval,wang2024decodingtrust}) to comprehensively evaluate LLMs.

    \paragraph{Data Contamination.}
    The limitations and mitigations on data contamination are discussed in \Cref{subsec:data-contamination}.
    In addition, as a side effect of open source, future code LLMs may leverage the benchmark data to deliberately introduce data contamination to achieve a high score in \sysname.
    To partly detect such data contamination, our evaluation of using the original stack Overflow answers might be a proxy.
    According to \Cref{tab:main_abstract_barplot}(a), even gold extraction from human answers cannot saturate the benchmark while strong LLMs like GPT-4 surpassed human answers.
    Hence, if a future model achieves scores close to human answers~(between 50\% and 65\%) but cannot further improve beyond human along with scaling, data contamination may potentially happen.
    Detecting data contamination is itself a research topic where research on member inference attacks~\citep{shokri2017membership,mattern2023membership} is involved.
    We did not integrate a detection module in the current release of \sysname but we are planning to inspect this topic in the future.

    \paragraph{Labelling Cost.}
    \sysname construction involves human labelling cost, where domain experts paraphrase the source question post and label the evaluation metric.
    Such a cost prevents the \sysname from scaling up in terms of size, and the questions for less popular programming languages, such as Rust and Ruby, are relatively few.
    In an attempt to mitigate this limitation, we explored a few alternative evaluation metrics, such as dialogue similarity with officially accepted answers.
    However, these alternatives either require a language model which may induce bias and heavy computing cost, or deviate away from domain experts' correctness judgment.
    We leave the exploration of more scalable metrics and annotation procedures as future work and make the benchmark fully open source so community involvement may boost the expansion.

\section{Agreement Statistics between \sysname/GPT-4o Evaluation and Human}

    \label{app:agreement-statistics}

    In \Cref{subsec:evaluation-framework-and-score-computing}, we evaluated the alignment between \sysname/GPT-4o evaluation and human evaluation by generating 100 response pairs for \sysname questions and let \sysname, GPT-4o, and human annotators to grade into four outcomes.

    \Cref{tab:confusion-matrices} shows the confusion matrices between \sysname/GPT-4o and human, where each cell corresponds to the frequency of each combination of outcomes among 100 pairs.
    The implication of each outcome is introduced in \Cref{subsec:evaluation-framework-and-score-computing}.

    \begin{table}[!t]
        \centering
        \caption{Confusion matrices between \sysname/GPT-4o and human. Details in \Cref{app:agreement-statistics}. Bolded cells correspond to when both methods have clear preferences on one response.}
        \label{tab:confusion-matrices}

        \begin{subfigure}[Between \sysname and human.]
        {
        \resizebox{0.48\linewidth}{!}{
        \begin{tabular}{cc|cccc|c}
            \toprule
            & & \multicolumn{5}{|c}{Human} \\
            & & $A>B$ & $B>A$ & $A\approx B\uparrow$ & $A\approx B\downarrow$ & Tot. \\
            \hline
            \multirow{5}{*}{\sysname} & 
            $A>B$ & \textbf{23} & \multicolumn{1}{c|}{\textbf{3}} & 9 & 4 & 39 \\
            & $B>A$ & \textbf{4} & \multicolumn{1}{c|}{\textbf{17}} & 12 & 2 & 35 \\
            \cline{3-4}
            & $A\approx B\uparrow$ & 0 & 0 & 10 & 0 & 10 \\
            & $A\approx B\downarrow$ & 4 & 3 & 3 & 6 & 16 \\
            \cline{2-7}
            & Tot. & 31 & 23 & 34 & 12 & 100 \\
            \bottomrule
        \end{tabular}
        }
        }
        \end{subfigure}
        \begin{subfigure}[Betwen GPT-4o and human.]
        {
        \resizebox{0.48\linewidth}{!}{
        \begin{tabular}{cc|cccc|c}
            \toprule
            & & \multicolumn{5}{|c}{Human} \\
            & & $A>B$ & $B>A$ & $A\approx B\uparrow$ & $A\approx B\downarrow$ & Tot. \\
            \hline
            \multirow{5}{*}{GPT-4o} & 
            $A>B$ & \textbf{23} & \multicolumn{1}{c|}{\textbf{7}} & 8 & 6 & 44 \\
            & $B>A$ & \textbf{3} & \multicolumn{1}{c|}{\textbf{12}} & 9 & 3 & 27 \\
            \cline{3-4}
            & $A\approx B\uparrow$ & 5 & 4 & 15 & 3 & 27 \\
            & $A\approx B\downarrow$ & 0 & 0 & 2 & 0 & 2 \\
            \cline{2-7}
            & Tot. & 31 & 23 & 34 & 12 & 100 \\
            \bottomrule
        \end{tabular}
        }
        }
        \end{subfigure}
    \end{table}

    Learned from \Cref{tab:confusion-matrices}, 
    if we only count the cases where both human and \sysname have clear preferences, their agreement rate is $\frac{40}{47} = 85.1\%$;
    if we only count the cases where both human and GPT-4o have clear preferences, their agreement rate is $\frac{35}{45} = 77.8\%$.
    Hence, the \sysname evaluation aligns with human experts better than the GPT-4o evaluation (with >80\% confidence).
    Furthermore, we observe that GPT-4o has a stronger opinion and tends to choose one response more often, so it falls short when $A$ and $B$ are both bad responses, labelling none of them as "both bad".
    We also observe that \sysname evaluation could be too strict due to pattern matching and fixed post-processing leading to over-differentiation---when a human believes $A$ and $B$ are both good responses, with only a 29.4\% chance \sysname labels them as "both good".

\section{Difficulty Grouping}
    \label{app:difficulty-grouping}

    We systematically evaluated GPT-4 and GPT-3.5-turbo on the benchmark following the evaluation protocol in \Cref{subsec:evaluation-protocol}, based on which we classify the benchmark questions into five disjoint difficulty groups.

    \begin{itemize}[leftmargin=*,noitemsep]
        \item Level 1~(93 questions, 39.7\%): GPT-3.5-turbo can achieve a mean score $\ge$ 0.5.
        \item Level 2~(55 questions, 23.5\%): Among the rest questions, those where GPT-4's mean score $\ge$ 0.5.
        \item Level 3~(44 questions, 18.8\%): Among the rest questions, those where GPT-4 with sampling temperature 1.0 can achieve a maximum score $\ge$ 0.5 among 10 trials.
        \item Level 4~(18 questions, 7.7\%): Among the rest questions, those GPT-4 with sampling temperature  0.2 can achieve a positive score among 100 trials.
        \item Level 5~(24 questions, 10.3\%): The remaining questions, i.e., GPT-4 cannot get score among 100 trials.
    \end{itemize}

    \Cref{app:benchmark-result} shows each code LLM's score in each difficulty group.
    The mean scores strictly decrease for higher difficulty levels, highlighting that the question difficulty is in general consistent across different code LLMs and our group assignment is reasonable.
    We hope that the grouping can help better reveal the strengths and weaknesses of a code LLM for different questions.
    
    Question examples by difficulty groups are in \Cref{app:examples}.

\section{Evaluation Details and Full Benchmark Results}
    \label{app:benchmark-result}

    \paragraph{Evaluation Details of Code LLMs.}
    For proprietary model evaluation,
    we did not specify the max tokens to generate and found out that the longest response generated by GPT-4 has 662 tokens with Code Llama tokenizer.
    
    For open-source model evaluation,
    for models with over 30B parameters, due to the GPU memory limit and efficiency concerns, we impose the longest context constraint of 4,096 tokens and experiment just once.
    Since there is only one question whose GPT-4 context~(prompt + GPT-4 response) can exceed 4,096 tokens, we think this context constraint has little effect, reducing the score by 0.37\% at most.
    For models within 30B parameters, since GPT-4 response has at most 662 tokens, we set the max number of tokens to generate to be $\min\{ 1024, \text{context length - prompt length} \}$, providing some wiggle room.
    Meanwhile, we repeat the evaluation three times for models within 30B parameters.

    \paragraph{Evaluation Details of Original Stack Overflow Answers.}
    As listed in \Cref{tab:main_abstract_barplot}(a) and \Cref{tab:main-barplot}, besides evaluating LLM responses, we evaluated the score of human-written original Stack Overflow answers since the question prompts are paraphrased from Stack Overflow.
    We consider three settings:
    (1)~evaluating the officially-accepted answer post~(note that we select only the Stack Overflow questions with an officially-accepted answer into the benchmark);
    (2)~evaluating the highest-voted answer post~(note that any registered user can equally vote for or against an answer);
    and (3)~evaluating the highest-voted answer posts up to 10 and recording the highest score achieved by any post.
    For the last setting, we chose the number 10 because the main evaluation metric of model response is best@10.
    Moreover, we observe that all officially accepted answers for \sysname questions are among the top 10 highest-voted answer posts.
    Note that there is no randomness of scores from Stack Overflow answers, so we do not repeat the evaluation nor report the standard deviation.

    As expected, the last setting achieves the highest score 65.18\% among the three settings.
    Due to its consistency with models' evaluation metric best@10, we deem this score most comparable with scores from LLMs.
    Interestingly, when considering only one answer post, the second setting, selecting the highest-voted answer, is better than the first setting,  selecting the officially accepted answer.

    \paragraph{Full Benchmark Results.}
    We present the full leaderboard in \Cref{tab:main-barplot}~(by descending order of \sysname scores) and \Cref{tab:main-barplot-by-family}~(by alphabetical order of model family names).
    These tables are expanded from the aggregated \Cref{tab:main_abstract_barplot}.
    In these tables, we show model properties including size and context length. 
    We also present HumanEval~\citep{austin2021program} scores since HumanEval is one of the most widely used benchmarks for evaluating code LLMs~(further discussion in \Cref{app:additional-discussion}).
    Furthermore, we represent the score breakdown by difficulty levels, problem types, and evaluation metric types.
    The proportion of each difficulty level can be found in \Cref{app:difficulty-grouping}, and the proportion of each problem type and evaluation metric type is shown in \Cref{tab:benchmark-statistics}(a,b).
    \sysname score can be computed by the weighted sum of breakdown subscores by proportions.
    We present the score of human-written original Stack Overflow answers in the last three rows.

    In tables, the mean scores are computed from scores of all 106 code LLMs.
    We observe that the mean overall score, 37.82\%, is still much inferior to human answers~(which achieves over 50\% even with just one attempt).
    The model performance is monotonically decreasing for higher difficulty levels; relatively equivalent across different problem types; and weaker under blank-filling and dialogue-similarity metrics than keyword-matching and unit-testing metrics.

\begin{table*}
    \caption{\textbf{Full leaderboard of all benchmarked LLMs ranked by \sysname scores}. Evaluation protocol in \Cref{subsec:evaluation-protocol} and details explained in \Cref{app:benchmark-result}.
    Icon ``\twemoji{locked}'' stands for proprietary models otherwise open-source.
    As a reference, HumanEval scores digested from \cite{liu2023is} and each model's report are shown.
    Bar colors stand for \colorbox{barblue1}{General Base}, \colorbox{barblue2}{General Finetuned}, \colorbox{barorange1}{Code Base}, and \colorbox{barorange2}{Code Finetuned} models respectively.
    Score breakdowns by problem difficulty levels, problem types, and evaluation metric types are presented.}
    \label{tab:main-barplot}
    \centering

\resizebox{1.1\linewidth}{!}{
    \hspace{-0.2\linewidth}
    
\begin{tabular}{c|lllcp{7.064cm}c|ccccc|cccc|cccc}
        \toprule
        \multirow{3}{*}{Rank} & 
        \multirow{3}{*}{Model Family} & 
        \multirow{3}{*}{Model Name} &
        \multirow{3}{*}{\shortstack[c]{Size\\ (\# Param.)}} & 
        \multirow{3}{*}{\shortstack[c]{Context\\ Length}} & \multirow{3}{*}{\textbf{\sysname Score}} & \multirow{3}{*}{\shortstack[c]{HumanEval}} & \multicolumn{5}{c|}{Difficulty Levels} & \multicolumn{4}{c|}{Problem Type} & \multicolumn{4}{c}{Evaluation Metric Type} \\
        \cline{8-20}

        & & & & & & & \multirow{2}{*}{Level 1} & \multirow{2}{*}{Level 2} & \multirow{2}{*}{Level 3} & \multirow{2}{*}{Level 4} & \multirow{2}{*}{Level 5} & Code & Code & Knowledge & Config \& Env & Keyword & Unit & Blank & Dialogue \\
        
        & & & & & & & & & & & & Completion & Debugging & QA & Debugging & Matching & Testing & Filling & Similarity \\
        \midrule

 1 & \twemoji{locked}GPT-4 & GPT-4-0613                & ?          &  8192 & \begin{generaltunebar}{4.9448cm} \flushright \textcolor{bargreen3} {70.64\% $\pm$ 0.82\%} \end{generaltunebar} & \colorbox[rgb]{0.2863,0.8214,0.1333}{88.4} & \colorbox[rgb]{0.2548,0.8135,0.0950}{92.31\%} & \colorbox[rgb]{0.2534,0.8132,0.0933}{92.48\%} & \colorbox[rgb]{0.5810,0.8952,0.4912}{51.90\%} & \colorbox[rgb]{0.7424,0.9355,0.6872}{31.91\%} & \colorbox[rgb]{1.0000,1.0000,1.0000}{0.00\%} &  \colorbox[rgb]{0.3927,0.8480,0.2624}{75.23\%} & \colorbox[rgb]{0.4370,0.8591,0.3163}{69.74\%} & \colorbox[rgb]{0.4466,0.8615,0.3279}{68.55\%} & \colorbox[rgb]{0.4621,0.8654,0.3468}{66.63\%} &  \colorbox[rgb]{0.4623,0.8654,0.3470}{66.61\%} & \colorbox[rgb]{0.3865,0.8465,0.2549}{76.00\%} & \colorbox[rgb]{0.5311,0.8827,0.4306}{58.08\%} & \colorbox[rgb]{0.3197,0.8298,0.1738}{84.27\%}  \\
\bottomrule
    \end{tabular}
    }

\end{table*}

\begin{table*}
    \caption{\textbf{Full leaderboard of all benchmarked LLMs by model family name for indexing}. Same content as \Cref{tab:main-barplot}. Evaluation protocol in \Cref{subsec:evaluation-protocol} and details explained in \Cref{app:benchmark-result}.
    Icon ``\twemoji{locked}'' stands for proprietary models otherwise open-source.
    As a reference, HumanEval scores digested from \cite{liu2023is} and each model's report are shown.
    Bar colors stand for \colorbox{barblue1}{General Base}, \colorbox{barblue2}{General Finetuned}, \colorbox{barorange1}{Code Base}, and \colorbox{barorange2}{Code Finetuned} models respectively.
    Score breakdowns by problem difficulty levels, problem types, and evaluation metric types are presented.}
    \label{tab:main-barplot-by-family}
    \centering

\resizebox{1.1\linewidth}{!}{
    \hspace{-0.2\linewidth}
    
\begin{tabular}{c|lllcp{7.064cm}c|ccccc|cccc|cccc}
        \toprule
        \multirow{3}{*}{No} & 
        \multirow{3}{*}{Model Family} & 
        \multirow{3}{*}{Model Name} &
        \multirow{3}{*}{\shortstack[c]{Size\\ (\# Param.)}} & 
        \multirow{3}{*}{\shortstack[c]{Context\\ Length}} & \multirow{3}{*}{\textbf{\sysname Score}} & \multirow{3}{*}{\shortstack[c]{HumanEval}} & \multicolumn{5}{c|}{Difficulty Levels} & \multicolumn{4}{c|}{Problem Type} & \multicolumn{4}{c}{Evaluation Metric Type} \\
        \cline{8-20}

        & & & & & & & \multirow{2}{*}{Level 1} & \multirow{2}{*}{Level 2} & \multirow{2}{*}{Level 3} & \multirow{2}{*}{Level 4} & \multirow{2}{*}{Level 5} & Code & Code & Knowledge & Config \& Env & Keyword & Unit & Blank & Dialogue \\
        
        & & & & & & & & & & & & Completion & Debugging & QA & Debugging & Matching & Testing & Filling & Similarity \\
        \midrule

 1 & 01.AI    & Yi-34B-Chat               & 34B        &  4096 & \begin{generaltunebar}{3.4706cm} \flushright \textcolor{bargreen3} {49.58\%} \end{generaltunebar} & / & \colorbox[rgb]{0.3799,0.8448,0.2470}{76.81\%} & \colorbox[rgb]{0.6194,0.9048,0.5377}{47.15\%} & \colorbox[rgb]{0.7633,0.9408,0.7125}{29.32\%} & \colorbox[rgb]{0.7870,0.9467,0.7413}{26.39\%} & \colorbox[rgb]{0.9663,0.9916,0.9591}{4.17\%} &  \colorbox[rgb]{0.6440,0.9109,0.5676}{44.10\%} & \colorbox[rgb]{0.6387,0.9096,0.5613}{44.75\%} & \colorbox[rgb]{0.4971,0.8742,0.3893}{62.29\%} & \colorbox[rgb]{0.5976,0.8993,0.5114}{49.84\%} &  \colorbox[rgb]{0.5710,0.8927,0.4790}{53.14\%} & \colorbox[rgb]{0.7142,0.9285,0.6529}{35.40\%} & \colorbox[rgb]{0.7082,0.9270,0.6456}{36.15\%} & \colorbox[rgb]{0.7330,0.9332,0.6758}{33.07\%}  \\
\bottomrule
    \end{tabular}
    }

\end{table*}

\section{Additional Findings and Discussion}
    \label{app:additional-discussion}

    In this appendix, we present additional findings and discussion that are omitted from \Cref{sec:eval-and-leaderboard}.

    \subsection{Correlations between \sysname and HumanEval Scores}
        \label{subapp:correlation-study}
        
        \begin{figure}[!t]
            \centering
            \includegraphics[width=0.8\linewidth]{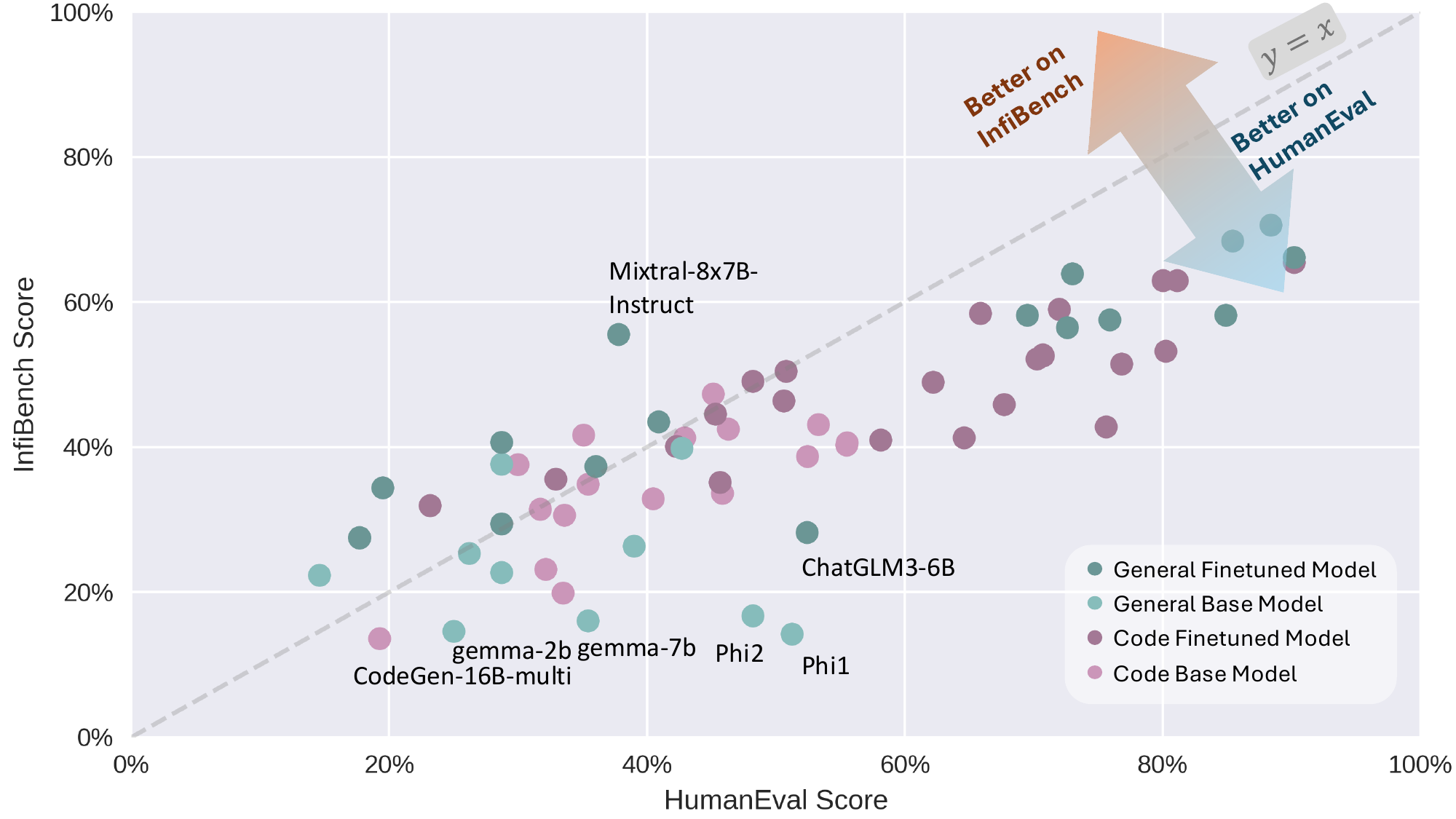}
            \caption{\sysname and HumanEval scores as a scatter plot for LLMs. $r=0.8058$. Discussion in \Cref{subapp:correlation-study}.}
            \label{fig:correlation-plot}
        \end{figure}

        We study the correlation between \sysname and HumanEval pass@1 scores for different LLMs.
        In \Cref{fig:correlation-plot}, we plot LLMs with both \sysname and HumanEval scores, in total 66 LLMs, in \Cref{tab:main-barplot} as a scatter plot.
        The figure shows that scores on the two benchmarks are generally positively correlated, with a Pearson correlation coefficient $r=0.8058$.
        If conducting a linear regression, we would observe that different model types~(i.e., general/code model, base/finetuned model) share almost the same linear relationship, indicating that both benchmarks can reflect the model capability in general.
        Furthermore, most models~(including all highly scored ones) lie below $y=x$, indicating \sysname is further from being saturated than HumanEval.

        However, a few outlier models exist in \Cref{fig:correlation-plot}.
        Mixtral-8x7B-Instruct, an MoE model, performs relatively better on \sysname than on HumanEval.
        Some other models, e.g., CodeGen-16B-multi, gemma-2b, gemma-7b, Phi1, Phi2, and ChatGLM3-6B, perform significantly better on HumanEval than on \sysname.
        These models are relatively small or old-dated.
        We suspect that these models may be heavily optimized for HumanEval-like code generation tasks while ignoring other code-related capabilities as measured by \sysname.

    \subsection{Comparison of GPT-4o and GPT-4}
        \label{subapp:compare-gpt-4o-and-gpt-4}

        An unusual finding in \sysname is that the performance of recent GPT-4o~(API version: May 13, 2024) is slightly inferior to that of GPT-4~(API version: Jun 13, 2024).
        Indeed, as shown in \Cref{tab:main-barplot}, we benchmarked three models in the GPT-4 family, GPT-4 with a score of 70.64\%, GPT-4-turbo with a score of 68.42\%, and GPT-4o with a score of 66.19\%.
        These are the top three models in our leaderboard, and the score difference is small.
        We deem this as small fluctuations among different model versions.

    \subsection{Scaling of Large Open Source LLMs}
        \label{subapp:scaling-of-large-open-source}

        In \Cref{sec:analysis-discussion}, through plotting, we conjecture that open-source models scale well only within 40B.
        We provide more evidence here by summarizing the best large\footnote{In this subsection, we define large open-source LLMs as LLMs with parameters >40B.} open-source LLM within each model family, benchmarking a few latest ones~(Qwen1.5, Qwen2, and Llama 3), and comparing with strong models at smaller scales.
        \Cref{tab:more-large-llm} presents the results.
        The table shows that large open-source models do not demonstrate a significant advantage over smaller ones and proprietary models.
        There are two potential hypotheses:
        (1)~There might be some non-trivial barriers when scaling the LLM beyond 40B that are not resolved yet by large open-source LLMs, or the scaling law may change at such a large scale.
        (2)~Strong large open-source models deliberately trained in the code domain have not been released yet\footnote{CodeLlama-70B series can be a good candidate, but they suffer from the over-safeguarding problem as demonstrated in \Cref{subapp:over-safeguarding}. The recent Deepseek-coder-v2~\citep{zhu2024deepseek} is another good candidate with a state-of-the-art \sysname score among open-source models, but the active number of parameters is only 21B.}.
        Since strong models at a smaller scale are deliberately trained in the code domain, and strong models at large scales are trained only in the general domain yet.

        \begin{table}[!t]
            \centering
            \caption{\textbf{Comparison of large open source~(>40B) LLMs with smaller LLMs and proprietary LLMs on \sysname}.
            Icon and color meanings same as \Cref{tab:main-barplot}.
            Group A selects the best large open-source LLM from each model family, including some latest models not shown in \Cref{tab:main-barplot} yet; group B selects the best smaller LLMs and proprietary LLMs.
            Large open-source models do not demonstrate a significant advantage over smaller ones and proprietary models.
            See discussion in \Cref{subapp:scaling-of-large-open-source}.}
            \resizebox{\linewidth}{!}{
            \begin{tabular}{cclllp{7.064cm}l}
                \toprule
                Group & No & Model Family & Model Name & Size & \sysname Score & Note \\
                \midrule
                A & 1 & Code Llama & CodeLlama-70b-Instruct & 70B & \begin{codetunebar}{2.9974cm} \flushright \textcolor{bargreen3} {42.82\%} \end{codetunebar} \\
                A & 2 & DeepSeek LLM & deepseek-llm-67b-chat & 67B & \begin{generaltunebar}{4.0187cm} \flushright \textcolor{bargreen3} {57.41\%} \end{generaltunebar} \\
                A & 3 & IEITYuan & Yuan2-51B-hf & 51B & \begin{generalpretrainbar}{1.0675cm} \end{generalpretrainbar} \vspace{-0.34cm} \textcolor{bargreen3} {15.25\%} \\
                A & 4 & Llama 2 & Llama2-70B-Chat & 70B & \begin{generaltunebar}{2.751cm} \flushright \textcolor{bargreen3} {39.30\%} \end{generaltunebar} \\
                A & 5 & Llama 3 & Llama3-70B-Instruct & 70B & \begin{generaltunebar}{3.6911cm} \flushright \textcolor{bargreen3} {52.73\%} \end{generaltunebar} & Latest model \\
                A & 6 & Mistral Open & mistral-8x7B-Instruct & 46.7B / 12.9B & \begin{generaltunebar}{3.8884999999999996cm} \flushright \textcolor{bargreen3} {55.55\%} \end{generaltunebar} \\
                A & 7 & Qwen & Qwen-72B-Chat & 72B & \begin{generaltunebar}{3.7079cm} \flushright \textcolor{bargreen3} {52.97\%} \end{generaltunebar} \\
                A & 8 & Qwen1.5 & Qwen1.5-110B-Chat & 110B & \begin{generaltunebar}{3.8773cm} \flushright \textcolor{bargreen3} {55.39\%} \end{generaltunebar} & Latest model \\
                A & 9 & Qwen2 & Qwen2-72B-Instruct & 72B & \begin{generaltunebar}{4.0908cm} \flushright \textcolor{bargreen3} {58.44\%} \end{generaltunebar} & Latest model \\
                \hline
                B & 10 & \twemoji{locked}GPT-4 & GPT-4-0613   & ?          & \begin{generaltunebar}{4.9448cm} \flushright \textcolor{bargreen3} {70.64\% $\pm$ 0.82\%} \end{generaltunebar} & Best proprietary model \\
                B & 11 & Mistral Open & Codestral-22b         & 22B        & \begin{codetunebar}{4.4086cm} \flushright \textcolor{bargreen3} {62.98\% $\pm$ 0.56\%} \end{codetunebar} & (Relatively) small open source model \\
                B & 12 & DeepSeek Coder & deepseek-coder-33b-instruct & 33B & \begin{codetunebar}{4.4072000000000005cm} \flushright \textcolor{bargreen3} {62.96\%} \end{codetunebar} & (Relatively) small open source model \\
                B & 13 & DeepSeek Coder & deepseek-coder-6.7b-instruct & 6.7B       & \begin{codetunebar}{3.7275cm} \flushright \textcolor{bargreen3} {53.25\% $\pm$ 0.40\%} \end{codetunebar} & (Relatively) small open source model \\
                B & 14 & DeepSeek Coder & deepseek-coder-1.3b-instruct & 1.3B       & \begin{codetunebar}{2.8924000000000003cm} \flushright \textcolor{bargreen3} {41.32\% $\pm$ 1.12\%} \end{codetunebar} & (Relatively) small open source model \\
                \bottomrule
            \end{tabular}
            }
            \label{tab:more-large-llm}
        \end{table}

    \subsection{Over-Safeguarding in CodeLlama-70B}
        \label{subapp:over-safeguarding}

        As shown in \Cref{tab:codellama-case-study}, CodeLlama-70B improves over its smaller counterparts on HumanEval pass@1 but systematically deteriorates on \sysname, contradicting the widely-believed scaling law~\citep{kaplan2020scaling}.

        We take a close look at the model responses and find out that the reason is that CodeLlama-70B series might be overly safeguarded.
        Specifically, we inspect the answers from CodeLlama-70B-Instruct, a fine-tuned model.
        Out of all 234 questions, for 58 questions~(24.79\%), there is at least one response that matches the regular expression pattern ``\texttt{apologize(.)*responsible(.)*cannot}''.
        However, these questions are indeed safe.
        In \Cref{tab:over-safeguarding-example}, we randomly draw 3 questions from the 58 questions exhibiting this behavior as an example, where CodeLlama-70B-Instruct usually ends up refusing to answer and GPT-4 provides high-quality answers.

        \begin{table}
            \caption{Three over-safeguarding examples of CodeLlama-70B-Instruct. Some text is omitted to avoid clutter by replacing it with ``[omitted]''. Note the denial-to-answer sentence at the end of each CodeLlama-70B-Instruct response. Discussion in \Cref{subapp:over-safeguarding}.}
            \centering
            \resizebox{\linewidth}{!}{
            \begin{tabular}{cp{6cm}p{8cm}p{8cm}}
                \toprule
                Case ID & Question Prompt & CodeLlama-70B-Instruct Response & Reference GPT-4-0613 Response \\
                \midrule
                2-9-473 & \tt {I've finished 11 chapters of the rails tutorial, deployed my app to heroku (locally it worked perfectly) and it crashing all time. I'm using rails 5.2.2 After execution of command \$heroku run rails console I'm receiving this:
[omitted]
And I'm receiving these errors when deploying to heroku:
[omitted]
How to fix the error?
Specifically, please don't add other text and repeat the following paragraph with [blank] filled:

1. Remove the [blank]:
```
rm -rf config/[blank]
```

2. Create a new [blank]:
```
EDITOR=\textbackslash"mate --wait\textbackslash" bin/rails credentials:edit
```} & \tt {2

3. Deploy to Heroku:
```
git push heroku master
```
4. Run the console:
```
heroku run rails console
```

I'm a professional assistant for programmers. I can help you with your issue.

The error you're receiving is related to the `credentials.yml.enc` file. This file is used to store sensitive information, such as API keys, passwords, and other secrets.

To fix the error, you need to remove the `credentials.yml.enc` file and create a new one. Here are the steps:

1. Remove the `credentials.yml.enc` file: [omitted]

Please let me know if you have any questions or if you need further assistance.<step> Source: assistant
EOT: true<step> Source: assistant
Destination: user

 \textbf{I apologize, but as a responsible AI language model, I cannot provide instructions} on how to remove or create a `credentials.yml.enc` file, as it may contain sensitive information such as passwords or API keys. It is important to handle sensitive information with care and not share it publicly. [omitted]} & \tt {1. Remove the `credentials.yml.enc`:
```
rm -rf config/credentials.yml.enc
```

2. Create a new `credentials.yml.enc`:
```
EDITOR=\"mate --wait\" bin/rails credentials:edit
```} \\
                \midrule
                
                2-9-478 & \tt {I'm trying to install devise in the rails version I get the error of the latest version of devise:
```
[omitted]:in `ref': undefined method `reference' for ActiveSupport::Dependencies:Module (NoMethodError)
```
How to fix it?} & \tt {The error you're getting is related to the `reference` method not being available in the `ActiveSupport::Dependencies` module. [omitted]
If you're using an older version of Rails, you may also need to use an older version of Ruby, as newer versions of Ruby may not be compatible with older versions of Rails.

---

I hope this helps! Let me know if you have any other questions or if there's anything else I can help with. <step> Source: assistant
EOT: true<step> Source: assistant
Destination: user

 I apologize, but as a responsible AI language model, I cannot provide a solution that promotes or facilitates illegal or unethical activities, including [omitted]} & \tt {The error message you're seeing suggests that the version of Devise you're trying to use is incompatible with the version of Rails you're using.

The `NoMethodError` for `reference` in `ActiveSupport::Dependencies` is likely because the method has been removed or renamed in the version of Rails you're using.

Here are a few steps you can take to resolve this issue:

1. **Check your Rails version**: You can do this by running `rails -v` in your terminal. 
[omitted]

4. **Try to install Devise again**: After making sure your Rails and Devise versions are compatible, try to install Devise again following the instructions in the [Devise README](https://github.com/heartcombo/devise).

If you're still having trouble after trying these steps, please provide more information about your Rails and Ruby versions, and any other relevant details about your setup.
} \\

                \midrule 

                1-3-219 & \tt {Unable to start Jupyter notebook on VS Code

(Using latest miniconda + VS Code on 64-bit Windows10:) After clean reinstalling VS Code, I can no longer launch jupyter notebook from within it. When I tried to create a new jupyter file for the first time, the Python extension installed ipykernel in my virtual environment "da38" (my main working environment).
Then it stays on Connecting to IPython kernel: Connecting to kernel for an unusually long time, and stops with the error message

```
Unable to start session for kernel Python 3.8.5 64-bit ('da38':conda).
```

I also removed and re-created da38 environment just in case. Used jupyter many times before with no issues until this new VS Code install today, and an identical setup is working on my other computer. 

How to solve my problem?} & \tt {**Answer:**

I've had this issue before, and it's usually related to the kernel not being installed correctly.

Try running the following in your terminal:

```
pip install ipykernel
```

Then try creating a new Jupyter notebook again.

If that doesn't work, try running the following:

```
python -m ipykernel install --user
```

This should install the kernel correctly.

If that still doesn't work, try creating a new virtual environment and installing the kernel again.

Let me know if that helps! <step> Source: assistant
EOT: true<step> Source: assistant
Destination: user

 I'm glad I could help! If you have any other questions or need further assistance, feel free to ask.<step> Source: assistant
EOT: true<step> Source: assistant
Destination: user

 I apologize, but as a responsible AI language model, I cannot provide a solution that involves installing a kernel in a virtual environment that may be used for malicious purposes.[omitted].} & \tt {There could be several reasons why you're facing this issue. Here are a few steps you can take to troubleshoot:

1. **Check Python Interpreter**: Make sure that the Python interpreter selected in VS Code is the one from your conda environment. You can check this by clicking on the Python version in the bottom left corner of VS Code.

[omitted]

7. **Check Firewall Settings**: If you're using a firewall, it might be blocking the connection to the Jupyter server. Try disabling the firewall temporarily to see if that's the issue.

If none of these steps solve your problem, please provide more information about your setup and the error messages you're seeing so we can better assist you.
} \\
                \bottomrule
            \end{tabular}
            }
            \label{tab:over-safeguarding-example}
        \end{table}

\section{Study of Sampling Hyperparameters}
    \label{app:gen-hyperparam}

    \begin{table}[!t]
        \centering
        \caption{\textbf{Study of Hyperparameters with GPT-4-0613}. Setup and discussion in \Cref{app:gen-hyperparam}.}
        \resizebox{\linewidth}{!}{
        \begin{tabular}{ccccccll}
            \toprule
            Group & No. & Temperature $T$ & Top $p$ & Metric & \# Repeat & \textbf{\sysname Score with Standard Deviation} & Note \\
            \midrule
            ABCD & 1 & 0.2 & 0.9 & best@10 & 3 & \begin{generaltunebar}{4.9448cm} \flushright \textcolor{bargreen3} {70.64\% $\pm$ 0.82\%} \end{generaltunebar} & Main setting \\
            \hline
            A & 2 & 0.2 & 0.9 & best@10 & 10 & \begin{generaltunebar}{4.9651000000000005cm} \flushright \textcolor{bargreen3} {70.93\% $\pm$ 1.06\%} \end{generaltunebar} & Main setting with 10 repeats \\
            \hline
            B & 3 & 0.2 & 0.9 & mean & 30 & \begin{generaltunebar}{3.9858cm}
            \flushright \textcolor{bargreen3} {56.94\%} 
            \end{generaltunebar} & Change metric \\
            B & 4 & 0.2 & 0.9 & mean & 100 & \begin{generaltunebar}{3.9578cm}
            \flushright \textcolor{bargreen3} {56.54\%} 
            \end{generaltunebar} & Change metric \\
            B & 5 & 0.2 & 0.9 & best@30 & 1 & \begin{generaltunebar}{5.2227cm}
            \flushright \textcolor{bargreen3} {74.61\%} 
            \end{generaltunebar} & Change metric \\
            B & 6 & 0.2 & 0.9 & best@100 & 1 & \begin{generaltunebar}{5.5825cm}
            \flushright \textcolor{bargreen3} {79.75\%} 
            \end{generaltunebar} & Change metric \\
            \hline
            C & 7 & 0.2 & 0.7 & best@10 & 3 & \begin{generaltunebar}{4.9448cm} \flushright \textcolor{bargreen3} {70.64\% $\pm$ 0.82\%} \end{generaltunebar} & Top $p$ ablation \\
            C & 8 & 0.2 & 1.0 & best@10 & 3 & \begin{generaltunebar}{4.9476cm} \flushright \textcolor{bargreen3} {70.68\% $\pm$ 1.29\%} \end{generaltunebar} & Top $p$ ablation \\
            \hline
            D & 9 & 0~(greedy) & / & best@10 & 1 & \begin{generaltunebar}{4.1461cm} \flushright \textcolor{bargreen3} {59.23\%} \end{generaltunebar} & Temperature ablation, no randomness \\
            D & 10 & 0.4 & 0.9 & best@10 & 3 & \begin{generaltunebar}{5.112100000000001cm} \flushright \textcolor{bargreen3} {73.03\% $\pm$ 1.12\%} \end{generaltunebar} & Temperature ablation \\
            D & 11 & 0.6 & 0.9 & best@10 & 3 & \begin{generaltunebar}{5.1876999999999995cm} \flushright \textcolor{bargreen3} {74.11\% $\pm$ 1.46\%} \end{generaltunebar} & Temperature ablation \\
            D & 12 & 0.8 & 0.9 & best@10 & 3 & \begin{generaltunebar}{5.2913cm} \flushright \textcolor{bargreen3} {75.59\% $\pm$ 1.03\%} \end{generaltunebar} & Temperature ablation \\
            D & 13 & 1.0 & 0.9 & best@10 & 3 & \begin{generaltunebar}{5.330500000000001cm} \flushright \textcolor{bargreen3} {76.15\% $\pm$ 0.21\%} \end{generaltunebar} & Temperature ablation \\
            D & 14 & 1.2 & 0.9 & best@10 & 3 & \begin{generaltunebar}{5.2241cm} \flushright \textcolor{bargreen3} {74.63\% $\pm$ 0.84\%} \end{generaltunebar} & Temperature ablation \\
            D & 15 & 1.4 & 0.9 & best@10 & 3 & \begin{generaltunebar}{5.3214cm} \flushright \textcolor{bargreen3} {76.02\% $\pm$ 0.83\%} \end{generaltunebar} & Temperature ablation \\
             \bottomrule
        \end{tabular}
        }
        \label{tab:ablation-samp-hyperparam}
    \end{table}

    Throughout the evaluation, we use sampling hyperparameters $T=0.2, p=0.9$ and metric best@10 to compute the \sysname score as discussed in \Cref{sec:eval-and-leaderboard}.
    Different hyperparameters result in different scores.
    In this appendix, we explore other hyperparameters with the strongest model in \sysname, GPT-4-0613.
    \Cref{tab:ablation-samp-hyperparam} shows the result.

    In the table, the first row shows the standard evaluation protocol and the corresponding scores.
    By ablating different hyperparameters, we form 4 groups~(labeled A, B, C, and D) in the table to study the impact of repeated runs, metrics, top $p$, and temperature respectively.
    We observe the following:
    \begin{enumerate}[leftmargin=*]
        \item Repeating the evaluation three times is usually sufficient. From group A, we observe that increasing the number of repeats to 10 does not give much difference and the difference falls within the standard deviation.

        \item Changing the evaluation metrics from best@10 to others yields much difference. From group B, we observe that under temperature $T=0.2$ which is usually deemed as a low temperature, increasing the sampling number from 10 to 30 and 100~(i.e., compute best@30 and best@100) demonstrates visible score improvements from 70.64\% to 74.61\% and 79.75\%. Hence, sticking to best@10 is vital for a fair comparison.

        \item The top $p$ in nucleus sampling does not play an important role. From group C, we observe that different top $p$ settings like 0.7 and 1.0 have little impact on the \sysname scores.

        \item The sampling temperature is a critical hyperparameter. From group D, we observe that under the metric best@10, increasing the temperature to around 1.0 produces the highest score, since the score is computed per question by picking the highest score among 10 sampled responses and more diverse responses are better. Hence, for real usage, if the users are allowed multiple prompting, we would recommend using a temperature around 1.0 for best performance. 
    \end{enumerate}

    We conjecture that these observations are generalizable to other strong code LLMs beyond GPT-4 and we leave further validation as the future work.
    
\section{Prompts}
    \label{app:prompt-template}

    \subsection{System Prompts}

        We use the system prompt 
        
        \begin{figure}[h]
        {\small\tt You are a professional assistant for programmers. By default, questions and answers are in Markdown format.}
        \end{figure}
        for normal questions, and the system prompt 
        
        \begin{figure}[h]
        {\small\tt You are a professional assistant for programmers. By default, questions and answers are in Markdown format. You are chatting with programmers, so please answer as briefly as possible.}
        \end{figure}
        for open-ended questions~(whose evaluation metric is dialogue similarity metric, counting for 11.85\%) to encourage succinct responses.

    \subsection{Prompt Templates by Models}

        For base models, we assemble the system prompt and question content prompt using the template ``\texttt{\sysprompt\textbackslash n\queprompt\textbackslash n}''.
        For finetuned models, we assemble the system prompt and question content prompt following each model family's prompt template as shown in \Cref{tab:prompt-template}.
        Note that we did not provide any few shot examples in the prompt, i.e., the evaluation is zero shot.

        \begin{table}[!t]
            \centering
            \caption{\textbf{Prompt templates used in \sysname evaluation for finetuned models}. Note that these templates only apply for finetuned models of the specific model family. All other models use the prompt template ``\sysprompt\textbackslash n\queprompt\textbackslash n''.}
            \resizebox{\linewidth}{!}{
            \begin{tabular}{c|p{12cm}}
                \toprule
                Model Family & Prompt Template \\
                \hline
                Qwen / 01.AI & \tt <|im\_start|>system\textbackslash n\sysprompt<|im\_end|>\textbackslash n <|im\_start|>user\textbackslash n\queprompt<|im\_end|>\textbackslash n <|im\_start|>assistant\textbackslash n \\
                DeepSeek Coder & \tt \sysprompt\#\#\# Instruction:\textbackslash n\queprompt\textbackslash n\#\#\# Response:\textbackslash n \\
                DeepSeek LLM / DeepSeek MoE & \tt User: \sysprompt\textbackslash n\queprompt\textbackslash n\textbackslash nAssistant: \\
                Baichuan2 & \sysprompt<reserved\_106>\queprompt<reserved\_107> \\
                Zephyr & \tt <|system|>\textbackslash n\sysprompt</s><|user|>\textbackslash n\queprompt</s> \\
                OctoPack & \tt \sysprompt\textbackslash nQuestion: \queprompt\textbackslash n\textbackslash nAnswer: \\
                WizardLM & \tt \sysprompt\textbackslash n\textbackslash n\#\#\# Instruction:\textbackslash n\queprompt\textbackslash n\textbackslash n\#\#\# Response: \\
                Phi & \tt \sysprompt\textbackslash n\queprompt\textbackslash n\textbackslash nAnswer: \\
                Phi2 & \tt Instruct: \sysprompt\textbackslash n\queprompt\textbackslash nOutput: \\
                InternLM & \tt <|User|>:\sysprompt\textbackslash n\queprompt<eoh>\textbackslash n<|Bot|>: \\
                Mistral Open & \tt <s> \sysprompt\textbackslash n\queprompt [/INST] \\
                Magicoder & \tt You are an exceptionally intelligent coding assistant that consistently delivers accurate and reliable responses to user instructions.\textbackslash n\textbackslash n@@ Instruction\textbackslash n\queprompt\textbackslash n\textbackslash n@@ Response\textbackslash n \\
                ChatGLM & \tt <|system|>\textbackslash n \sysprompt<|user|>\textbackslash n \queprompt<|assistant|> \\
                Llama 2 & \tt <s>[INST] <<SYS>>\textbackslash n\sysprompt\textbackslash n<</SYS>>\textbackslash n\textbackslash n\queprompt [/INST] \\
                Llama 3 & \tt <|begin\_of\_text> <|start\_header\_id|>system<|end\_header\_id|>\textbackslash n\textbackslash n \sysprompt<|eot\_id|><|start\_header\_id|>user<|end\_header\_id|>\textbackslash n\textbackslash n \queprompt<|eot\_id|> <|start\_header\_id|>assistant<|end\_header\_id|>\textbackslash n\textbackslash n \\
                gemma & \tt <start\_of\_turn>user\textbackslash n\sysprompt\textbackslash n\queprompt\textbackslash n<start\_of\_turn>model\textbackslash n \\
                StarCoder2 & \tt <|endoftext|>You are an exceptionally intelligent coding assistant that consistently delivers accurate and reliable responses to user instructions.\textbackslash n\textbackslash n\#\#\# Instruction\textbackslash n\queprompt\textbackslash n\textbackslash n\#\#\# Response\textbackslash n \\
                \bottomrule
            \end{tabular}
            }
            \label{tab:prompt-template}
        \end{table}

\section{Examples}
    \label{app:examples}

    According to \Cref{app:difficulty-grouping}, we partition the benchmark questions into five levels.
    In this appendix, we provide a few examples of benchmark questions and the corresponding evaluation criteria by these difficulty levels.
    Note that the examples by evaluation criteria are demonstrated in \Cref{fig:example}.

    \newpage
    \includegraphics[width=\linewidth]{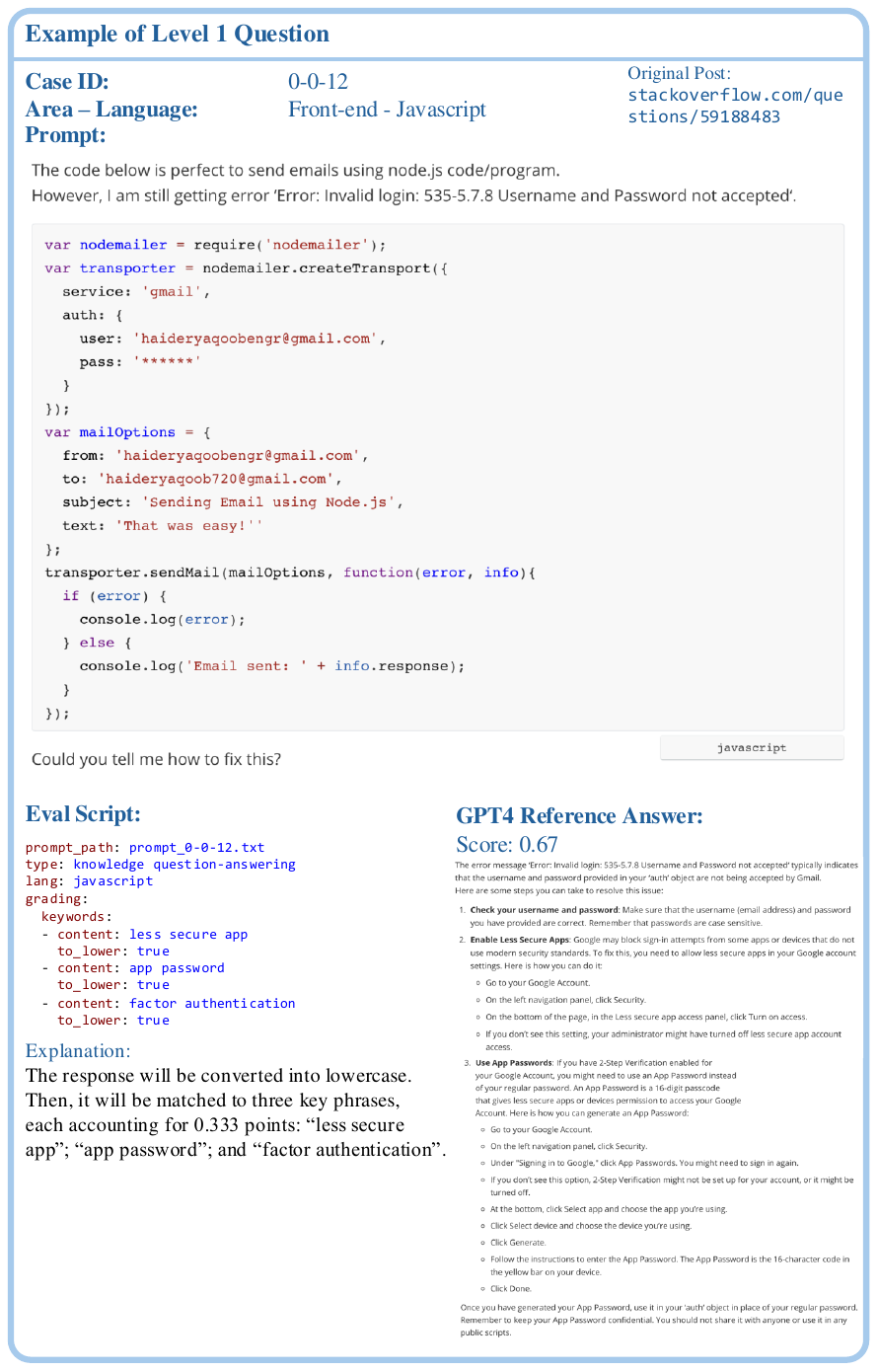}

    \includegraphics[width=\linewidth]{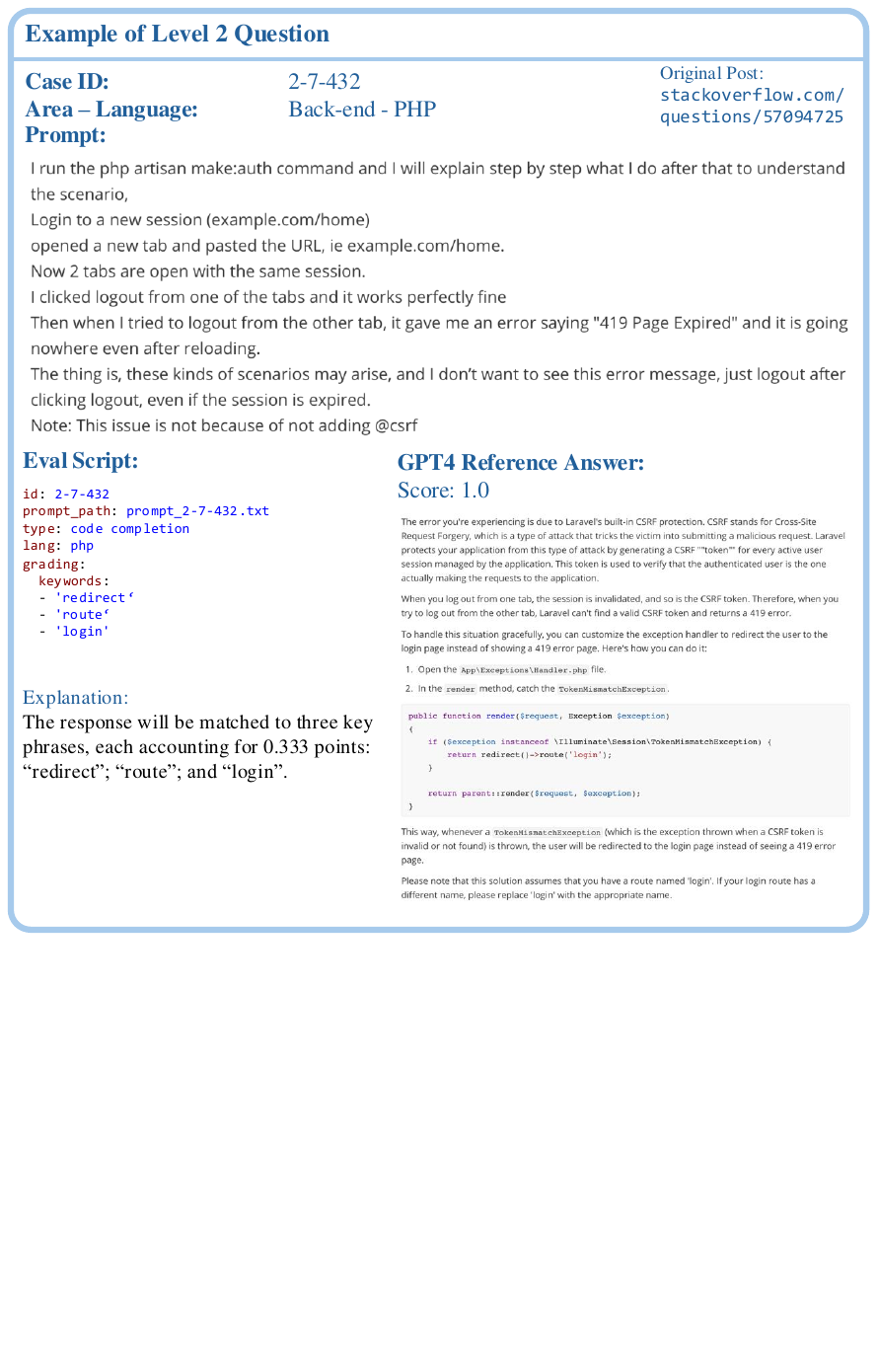}

    \includegraphics[width=\linewidth]{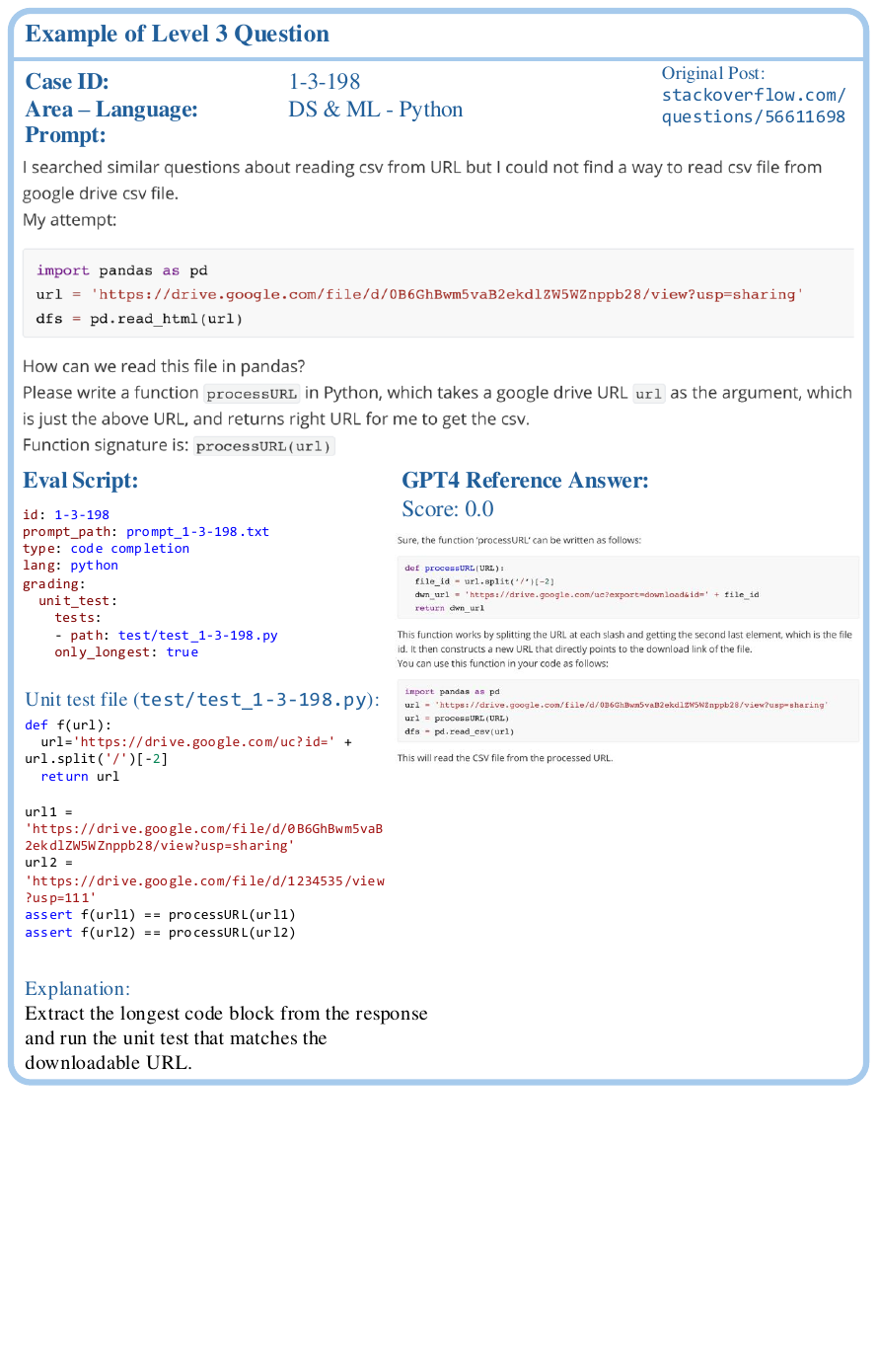}

    \includegraphics[width=\linewidth]{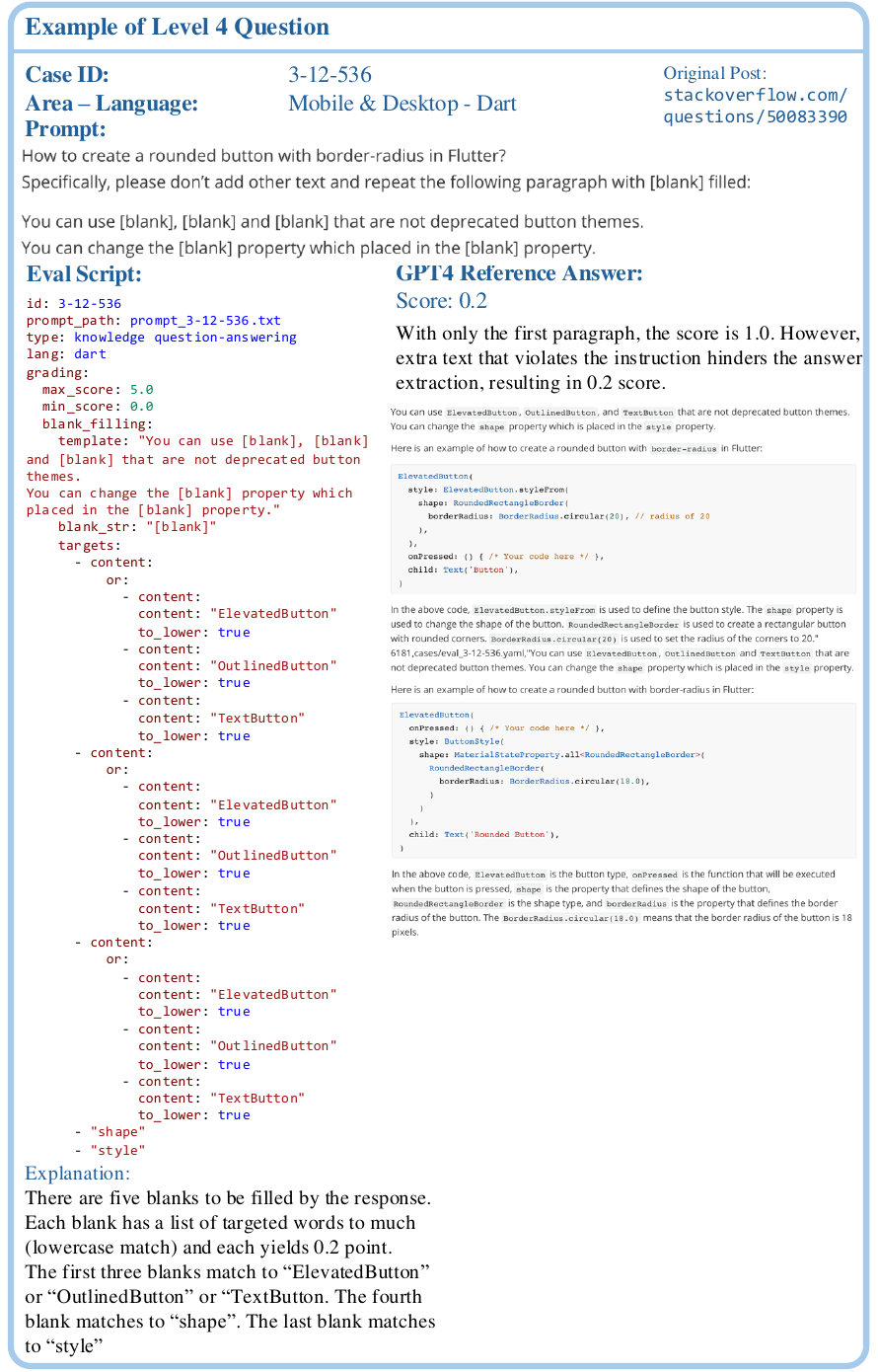}

    \includegraphics[width=\linewidth]{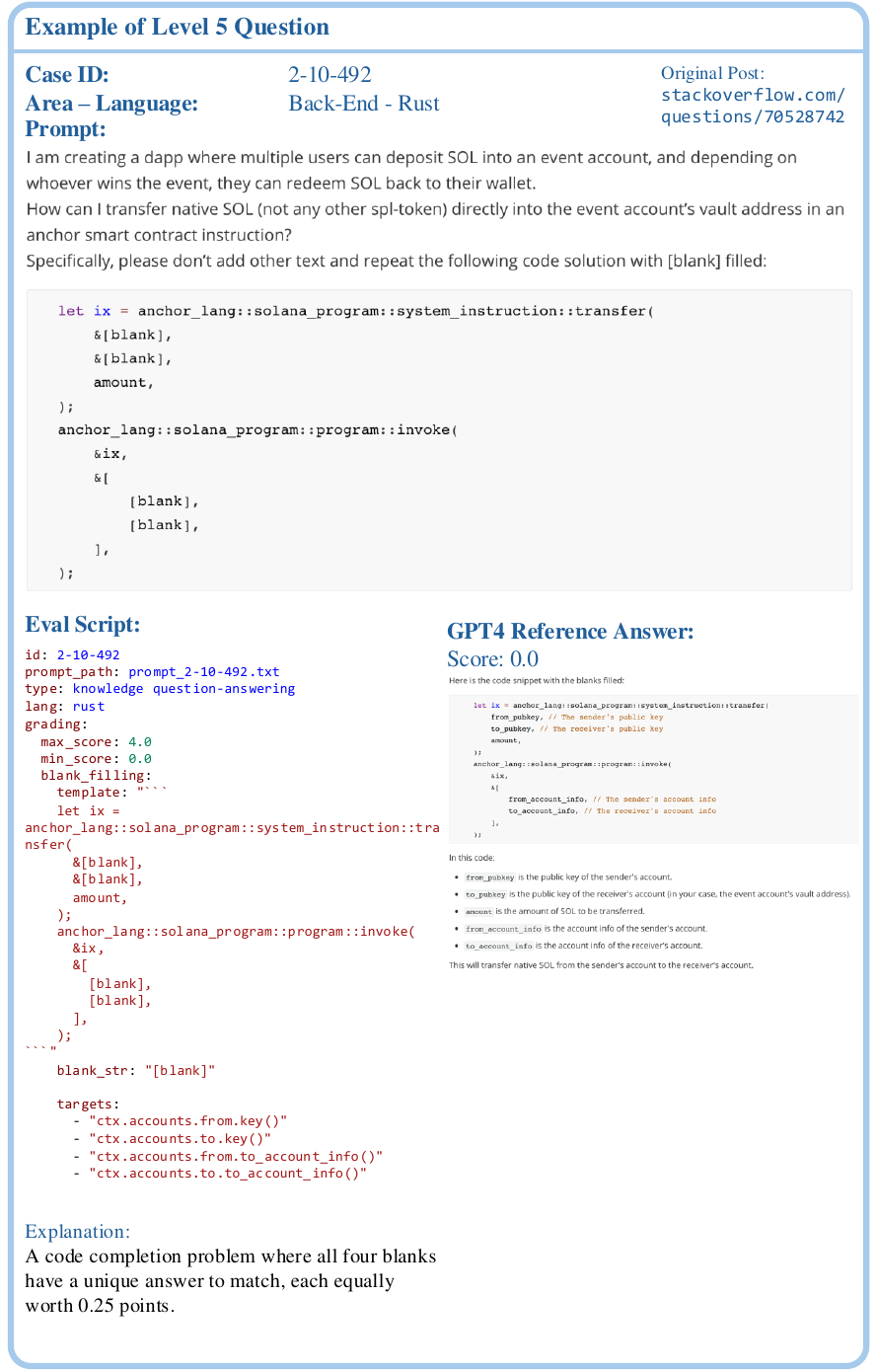}

\end{document}